\documentclass[a4paper,12pt]{article}

\usepackage{cite}
\usepackage{pstricks}
\usepackage{multirow}
\usepackage{latexsym}
\usepackage{bm}
\usepackage{epstopdf}

\usepackage{graphicx}

\usepackage{epsfig}
\usepackage{color}
\usepackage{tabularx}
\usepackage{subfig}
\usepackage{amsmath,amssymb}

\begin{document}
\vspace*{-0.2in}
\begin{flushright}
OSU-HEP-12-07\\
\end{flushright}

\vspace*{0.5cm}

\begin{center}
{\large\bf Higgs Boson of Mass 125 GeV in GMSB Models\\[0.05in]
 with Messenger--Matter Mixing}\\
\end{center}

\vspace{0.4cm}
\begin{center}
{\bf A. Albaid}$^{a,b,}$\footnote{Email: abdelhamid.albaid@wichita.edu} and {\bf K.S. Babu}$^{a,}$\footnote{E-mail: babu@okstate.edu}

\vspace*{0.2cm}

{\em $^{a}$Department of Physics, Oklahoma State University, Stillwater, OK 74078, USA }
{\em $^{b}$Department of Mathematics, Statistics, and Physics, 1845 Fairmount, Box 33,\\
 Wichita State University, Wichita, KS 67260-0033, USA}
\vspace*{0.1in}
\end{center}

\begin{abstract}

We investigate the effects of messenger--matter mixing on the lightest CP--even Higgs boson mass $m_h$
in gauge--mediated supersymmetry
breaking models. It is shown that with such mixings $m_h$ can be raised to about
125 GeV, even when the superparticles have sub--TeV masses, and when the
gravitino has a cosmologically preferred sub--keV mass.  In minimal gauge mediation
without messenger--matter mixing, realizing $m_h \sim 125$ GeV would require multi--TeV SUSY spectrum.
The increase in $m_h$ due to messenger--matter mixing is maximal in the case of messengers belonging to
$10+\overline{10}$ of $SU(5)$ unification, while it is still significant
when they belong to $5+\overline{5}$ of $SU(5)$.  Our results are compatible
with gauge coupling unification, perturbativity, and the unification of
messenger Yukawa couplings. We embed these models into a grand unification framework with
a $U(1)$ flavor symmetry that addresses the fermion mass hierarchy and
generates naturally large neutrino mixing angles. While SUSY mediated flavor changing
processes are sufficiently suppressed in such an embedding, small new contributions
to $K^0-\overline{K^0}$ mixing can resolve the apparent discrepancy in the CP asymmetry parameters $\sin2\beta$ and $\epsilon_K$.

\end{abstract}

\newpage

\renewcommand{\thefootnote}{\arabic{footnote}}\setcounter{footnote}{0}
\setlength{\baselineskip}{18pt}

\section{Introduction}

The Higgs boson continues to be a subject of intense scrutiny.  The CMS \cite{CMS}
and ATLAS \cite{ATLAS} collaborations have recently reported observation of a new particle
with a mass near 125 GeV with properties that are 
consistent with the Standard Model Higgs boson.
Each of these experiments has  a statistical significance of 5 standard deviations.  
The observed mass of the particle is consistent with exclusions  obtained
previously for the SM Higgs boson, viz., $m_h < 114.4$ GeV excluded by the LEP experiments \cite{LEP},
$156 ~{\rm GeV} < m_h < 177$ GeV excluded by the Tevatron experiments \cite{CDFD0},
$131 <m_h <237$ GeV and $251 < m_h< 453$ GeV excluded by ATLAS \cite{ATLAS1},
and $127 < m_h < 600$ GeV excluded by CMS \cite{CMS1}, all at
95\% CL. A light Higgs boson is a characteristic prediction of supersymmetric
theories.  In view of the interest in $m_h \sim 125$ GeV, in this paper we investigate
expectations for the lightest CP--even Higgs boson mass in a popular class of SUSY models, viz.,
gauge mediated supersymmetry breaking (GMSB) models \cite{GMSB,GR}.

Simple versions of GMSB models would predict $m_h < 118$ GeV, if the SUSY particle
masses lie below 2 TeV or so \cite{Suspect,ilia}.  In the general MSSM with arbitrary soft breaking, this mass can
be as large as about 130 GeV, with sparticle masses below 2 TeV.  GMSB models set a restriction
on the soft SUSY breaking trilinear $A_t$ parameter, $A_t = 0$ at the messenger scale, which disallows maximal
stop mixing, leading to
the reduced upper limit on $m_h$.  If larger values of the sparticle masses are allowed within GMSB,
the limit of 118 GeV can be raised somewhat, but masses in excess of 2 TeV, especially for the stops,
would go against naturalness in the Higgs mass, and would also render SUSY untestable
at the Large Hadron Collider.  Here we address the general question: How large can $m_h$ be in minimal
gauge mediation without making sparticles beyond the reach of LHC?  We find that $m_h$ can be raised naturally to about $(125-126)$ GeV,
with SUSY particles all below 2 TeV, if the messengers of SUSY breaking
are allowed to mix with the Standard Model fields.  Such a scenario would make GMSB models compatible with the recent Higgs  observations.

Along with the trilinear $A$--terms, the bilinear SUSY breaking $B$ term (${\cal L} \supset -\mu B H_u H_d$) also
vanishes at the messenger scale in a class of minimal GMSB models.
Upon renormalization, this condition would determine through the minimization of the Higgs potential
the value of the parameter $\tan\beta$, which turns out to
be typically large, $\tan\beta \approx (35-40)$ \cite{BKW,Borzumati,RS}. We show that much lower values, $\tan\beta \approx (2-8)$, can be realized
in presence of order one mixed messenger--matter Yukawa couplings.  Thus, the entire range $\tan\beta = (2-40)$ can be realized
with vanishing $B$ term at the messenger scale.
The SUSY spectroscopy of these models is different from those of GMSB without
messenger--matter mixing, and leads to relatively light stops. The mixing of messenger fields with the MSSM fields has a cosmological
advantage that it would break ``messenger number" that would have led to a stable messenger particle, which is not an ideal
candidate for dark matter.  While SUSY flavor violation arising from messenger--matter
mixing is not excessive, the proposed scenario predicts small but observable flavor effects. When the models presented are embedded
in a unified $SU(5)$ framework with a flavor $U(1)$ symmetry, it is found that the CP asymmetry parameter $\epsilon_K$
in the $K$ meson system is slightly modified, which can explain the apparent discrepancy in the extracted value of $\sin2\beta$ in the $B$ meson system.
We also find that the rare decay $\mu \rightarrow e\gamma$ should be accessible to the next
generation experiments.

This paper is organized as follows.  In Sec. 2 we summarize the salient features of minimal gauge mediation.  In Sec. 3
we discuss the upper limit on the lightest Higgs boson mass in GMSB models including messenger--matter mixing.  Two models are
studied, a $5+\overline{5}$ messenger model, and a $10+\overline{10}$ messenger model.  In this section we also discuss
the sparticle spectroscopy, allowing for messenger--matter mixing, and obtain limits on $\tan\beta$ with the boundary
condition $B=0$ at the messenger scale.  In Sec. 4 we discuss flavor violation
arising from messenger--matter mixing in the two models.  Here we embed these models in a grand unification framework
based on $SU(5)$ along with a flavor $U(1)$ symmetry that addresses the quark and lepton mass and mixing hierarchies.
Sec. 5 has our conclusions. The relevant renormalization group equations (RGE) for the two models are given in Appendix A, and the GMSB boundary
conditions for the mass parameters are derived in the Appendix B.  Preliminary results of this work were presented at PHENO 2011 \cite{Pheno}.

\section {Essential features of minimal gauge mediation}

GMSB models are well motivated, since SUSY solves the hierarchy problem, and gauge
mediation of SUSY breaking solves the SUSY flavor problem.  These models also predict correctly the unification
of the three gauge couplings, leading to an eventual embedding in a grand unified theory (GUT) such as $SU(5)$.
Gravity mediation of SUSY breaking (SUGRA) also shares these features, except that generically it would lead to
excessive flavor changing processes mediated by the SUSY particles.  Consistency of SUGRA models with experiments
would typically require two assumptions \cite{mSUGRA}:
(i) the soft masses of sparticles in any given sector are universal, and (ii) the trilinear $A$--terms
are proportional to the corresponding Yukawa couplings.  Such assumptions are not necessary in
GMSB models, rather they are automatic consequences.  GMSB models assume that SUSY is dynamically
broken in a secluded sector, and that this breaking is communicated to the MSSM sector via the SM gauge interactions
by a set of messenger fields which are charged under the SM.
Owing to the universality of the gauge interactions the soft SUSY breaking mass parameters would be flavor universal,
and the induced $A$--terms would be proportional to the Yukawa couplings.

Minimal gauge mediation assumes that a gauge singlet superfield $Z$ develops nonzero vacuum expectation values (VEVs)
along its scalar component $\left\langle Z \right \rangle$ as well as along its auxiliary component
$\left\langle F_Z \right \rangle$.  This field couples to a set of messenger fields $\Phi_i$ and $\overline{\Phi}_i$
which transform vectorially under the SM gauge symmetry:
\begin{equation}
W = \lambda_i Z \Phi_i \overline{\Phi}_i~.
\label{sup}
\end{equation}
$\left\langle F_Z \right \rangle \neq 0$ would split the masses of the scalars in $\Phi_i$ from the corresponding
fermions.  This breaking of SUSY is communicated to the SM sector via loops involving the SM gauge bosons.  The gaugino masses
and the scalar masses for the MSSM fields at the messenger scale are given by
\begin{eqnarray}
M_a &=& \frac{\alpha_a}{4 \pi} \Lambda \,n_a(i)\, g(x_i) ~~~~(a=1-3), \nonumber \\
\tilde{m}^2 &=& 2 \Lambda^2 \sum_{a=1}^3 \left(\frac{\alpha_a}{4\pi}\right)^2C_a n_a(i)f(x_i)~.
\label{spectrum}
\end{eqnarray}
Here $\Lambda \equiv \left\langle F_Z \right\rangle/\left\langle Z \right\rangle$ and $n_a(i)$ is the Dynkin index
of the messenger pair $\Phi_i$ with $n_a(i) = 1$ for $N+\overline{N}$ of $SU(N)$.  $C_a$ is the quadratic
Casimir invariant of the relevant MSSM scalar with $C_a = (N^2-1)/(2N)$ for $N$--plet of $SU(N)$ and $C_a = (3/5)\, Y^2$ for $U(1)_Y$.
The functions $f(x_i)$ and $g(x_i)$ can be found, for e.g.,  in Ref. \cite{kribs}, and are nearly equal to one for small values of $x_i$,
defined as $x_i = |\langle F_Z\rangle /\lambda_i \langle Z \rangle^2|$ with $x_i < 1$ necessary for color and
charge conservation.  In addition, GMSB models impose the following boundary conditions at the messenger scale on the MSSM trilinear and bilinear soft SUSY breaking parameters:
\begin{eqnarray}
A_f &=& 0 ~~{\rm for ~ all}~ f
\nonumber \\
B &=& 0~.
\label{AB}
\end{eqnarray}
The second of these relations, $B=0$, is sometimes ignored anticipating some mechanism that explains the
magnitude of the $\mu$ parameter \cite{dimo,bm}.  For example, in Ref. \cite{bm}, a flavor symmetry is assumed in
the singlet ($Z$) sector, so that $B\mu \ll \mu^2$ or $B\mu \sim \mu^2$ can be realized at the messenger scale ($M_{\rm mess}$),
depending on the assignment of flavor charges.  In our analysis we shall allow for $B=0$ as well as $B\neq 0$ at
$M_{\rm mess}$.

A few features are worth emphasizing on Eqs. (\ref{spectrum}) and (\ref{AB}).  Sparticles of
a given quantum number are all degenerate in mass, which is crucial in solving the SUSY flavor problem.  The induced
trilinear couplings would be proportional to the respective Yukawa couplings owing to the vanishing of the
$A$ terms, also crucial for solving the SUSY flavor problem.  The minimal GMSB models also have only a small number
of effective parameters.

The gravitino is the lightest supersymmetric particles in minimal GMSB.  Its mass is given by $m_{3/2} = \langle F_Z \rangle/(\sqrt{3}\,k \,M_{\rm Pl})$ where $M_{\rm Pl}$ is the reduced Planck mass, and $k$ is a typical Yukawa coupling of the type $\lambda$ given in Eq. (\ref{sup}). The cosmological
requirement that the gravitinos do not overclose the universe requires $m_{3/2} < $ keV \cite{primack}, which in turn requires
$\sqrt{\langle F_Z \rangle}
< \sqrt{k}\,\, 2 \cdot 10^6$ GeV.  In GMSB with a single set of messenger fields, $M_{\rm mess} = \lambda \langle Z \rangle$, so that this
constraint would require $M_{\rm mess} = \lambda \langle F_Z\rangle/\Lambda$ to obey $M_{\rm mess} \leq \lambda^2 \,\,(10^8$ GeV), where $k =\lambda$ and
$\Lambda = 3 \cdot 10^4$ GeV, its lowest allowed value, are used.  Perturbativity would require $\lambda < 1$,
so that cosmology prefers $M_{\rm mess} < 10^8$ GeV.  Since there are ways around the gravitino overclosure
problem, such as by late decays of particles, or other ways of entropy dumping, the cosmological limit is not absolute.
In our analysis we find fully consistent solutions when this limit is satisfied. We also allow $M_{\rm mess}$ to be greater
than $10^8$ GeV, as large as $10^{14}$ GeV.
Any larger value would lead to $m_{3/2} > 1$ GeV, and thus  generate supergravity contributions to the scalar masses that can bring back the SUSY flavor problem.

The messenger fields, which are taken to be vector--like under the SM gauge symmetry, are usually assumed to form complete multiplets
of a grand unified group.  This is motivated by the observed meeting of the three gauge couplings at a scale near $2 \cdot 10^{16}$ GeV
when extrapolated with the MSSM spectrum.
Complete multiplets of a GUT symmetry group such as $SU(5)$ would preserve this
 successful unification (modulo small two--loop effects).  Messenger fields belonging to $5+\overline{5}$ of $SU(5)$ or
 $10+\overline{10}$ of $SU(5)$ are then the simplest choices.  One could introduce multiple copies of these fields, or one
 could introduce both of them simultaneously.  We shall consider only two minimal choices in this paper, viz., having either
 one pair of $5+\overline{5}$ or one pair of $10+\overline{10}$ messenger fields.

Messenger fields belonging to $5+\overline{5}$ of $SU(5)$ or $10+\overline{10}$ of $SU(5)$ can mix with the MSSM superfileds.
If such mixings are written down arbitrarily, that would reintroduce SUSY flavor problem.  However, in the context of an underlying
flavor symmetry that addresses the mass and mixing hierarchy of quarks and leptons, it is not unreasonable to imagine
that significant mixing of the messenger fields occurs only with the third family fermions.  This is the situation
we investigate in the next sections.  Complete separation of messenger fields from the MSSM fields is in general
problematic for cosmology, since this would lead to messenger number conservation and a stable messenger particle,
which is not an ideal dark matter candidate \cite{dark}.  Messenger--matter mixing avoids this difficulty.
In presence of such mixings, the expressions given in Eqs. (\ref{spectrum})-(\ref{AB})
for the soft SUSY breaking parameters would receive
new contributions  \cite{dine2,GR2,Chacko}.  This can help increase the lightest Higgs boson mass of GMSB, and can lead to significantly different
SUSY spectrum.  We also point out that such mixings can modify the derived value of $\tan\beta$, which can now be quite
low, in the range of $2-8$, with order one Yukawa couplings.

\section{Higgs boson mass bound in GMSB models}

Low energy supersymmetry characteristically predicts one light Higgs boson. In
the MSSM, at the tree level, the lightest Higgs boson mass is bounded by $m_h \leq M_Z$.
Radiative corrections proportional to the top quark Yukawa couplings shift this
limit significantly \cite{higgs1,higgs2}.  Including the leading two loop corrections,
this mass can be written approximately as \cite{higgs2}
\begin{eqnarray}
m_h^2&=&M_Z^2\cos^22\beta\left(1-\frac{3}{8\pi^2}\frac{m_t^2}{v^2}t\right)\nonumber\\
&+&\frac{3}{4\pi^2}\frac{m_t^4}{v^2}\left[\frac{1}{2} X_t
+t+\frac{1}{16\pi^2}\left(\frac{3}{2}\frac{m_t^2}{v^2}-32\pi\alpha_3\right)\left(X_tt+t^2\right)\right],
\label{higgsmass}
\end{eqnarray}
where
\begin{eqnarray}
v^2=v_d^2+v^2_u,~~~t=\log\left(\frac{M_s^2}{M_t^2}\right),~~~
X_t=\frac{2\tilde{A}^2_t}{M_s^2}\left(1-\frac{\tilde{A}^2_t}{12M_s^2}\right),
\end{eqnarray}
with the scale $M_s$ defined in terms of the stop mass
eigenvalues as $M_s^2=\tilde{m}_{t_1}\tilde{m}_{t_2}$.  Here
$\tilde{A}_t=A_t-\mu\cot\beta$, with $A_t$ being the trilinear soft term for the stop.
Eq. (\ref{higgsmass}) is accurate to about 3 GeV, when compared with computational
packages such as SuSpect \cite{Suspect}
which do not make certain simplifying assumptions made in obtaining Eq. (\ref{higgsmass}).
Since we find that the numerical package SoftSusy consistently gives 2 GeV larger Higgs mass compared
to Eq. (\ref{higgsmass}), we find it appropriate to add 2 GeV to $m_h$ computed from Eq. (\ref{higgsmass})
for interpretation.
The upper bound on $m_h$ depends crucially on $M_s$ and the mixing parameter $X_t$.  It is maximal in
the case of maximal stop mixing (corresponding to $X_t = 6$), in which case $m_h = 130$ GeV can be realized
with all SUSY particles below 2 TeV.  The first boundary condition of Eq. (\ref{AB}) would
however forbid realizing maximal stop mixing in minimal GMSB.  The upper limit on $m_h$ in this case is
$m_h < 118$ GeV, with all sparticle masses below 2 TeV \cite{Suspect}.

In presence of messenger matter mixings, the boundary conditions Eqs. (\ref{spectrum})-(\ref{AB}) will receive
new contributions.  In such cases, near maximal mixing of the stops can be realized, as we show here, and
thus the upper limit on $m_h$ can be raised to about $(125-126)$ GeV.  New contributions to the $A$--terms also would
imply that the value of $\tan\beta$ derived with the condition $B=0$ at $M_{\rm mess}$ (this condition is un-altered
even with matter--messenger mixing) would be different.  Lower values of $\tan\beta$ are found, which can be
understood from the one--loop RGE for the $B$ parameter below $M_{\rm mess}$:
\begin{eqnarray}
\frac{dB}{dt}=\frac{1}{2\pi}(3\alpha_tA_t+3\alpha_2M_2+\frac{3}{5}\alpha_1M_1)\label{p4B}~,
\label{B-evolve}
\end{eqnarray}
where $\alpha_t=\frac{\lambda_t^2}{4\pi}$, $\lambda_t$ being the top quark
Yukawa coupling.
The non-zero initial value of $A_t$ modifies the evolution of $B$, which is related to $\tan\beta$ by the
electroweak symmetry breaking conditions given by
\begin{eqnarray}
\frac{M_Z^2}{2}&=&-|\mu|^2-\frac{m^2_{H_u}\tan^2\beta-m^2_{H_d}}{\tan^2\beta-1},\label{p4EW1}\\
\sin2\beta &=&\frac{2B\mu}{2|\mu|^2+m^2_{H_u}+m^2_{H_d}}\label{p4EW2}~.
\label{SSB}
\end{eqnarray}
The effect of non--zero $A_t$ is to decrease the value of $\tan\beta$.
For example in the
$10+\overline{10}$ messenger model, we find the range $1.6\leq\tan\beta\leq 7$ assuming $B=0$ at $M_{\rm mess}$
with order one messenger Yukawa couplings, corresponding to $10^{14}~ \rm{GeV}\geq M_{\rm{mess}}\geq 10^{5}~\rm{GeV}$.

\subsection{Higgs mass bound in the $5+\overline{5}$ messenger model}

\begin{figure}[h!]
\centering {\includegraphics[width=7.5 cm]{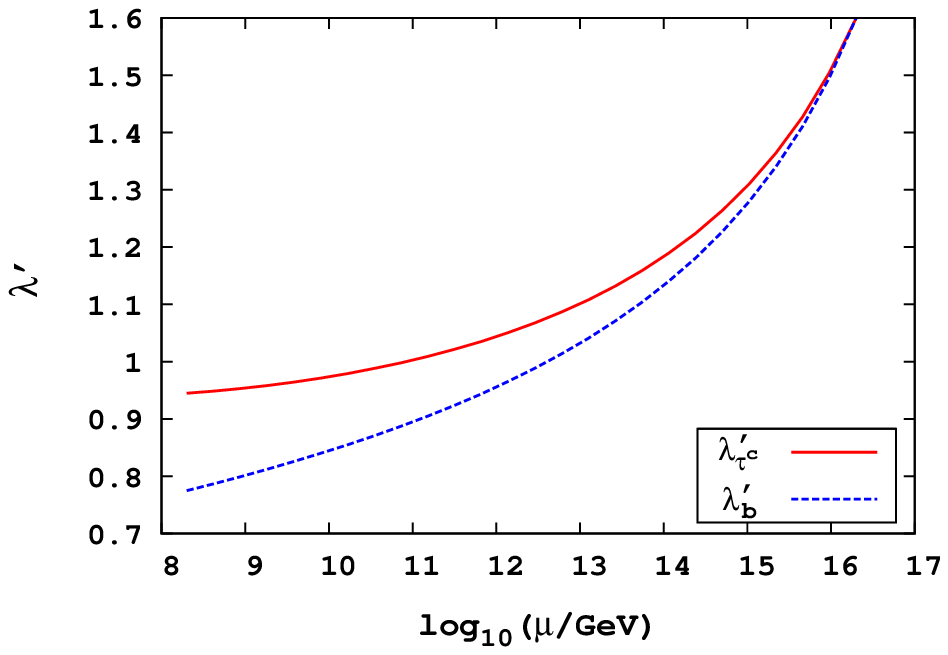}}
{\includegraphics[width=7.5 cm]{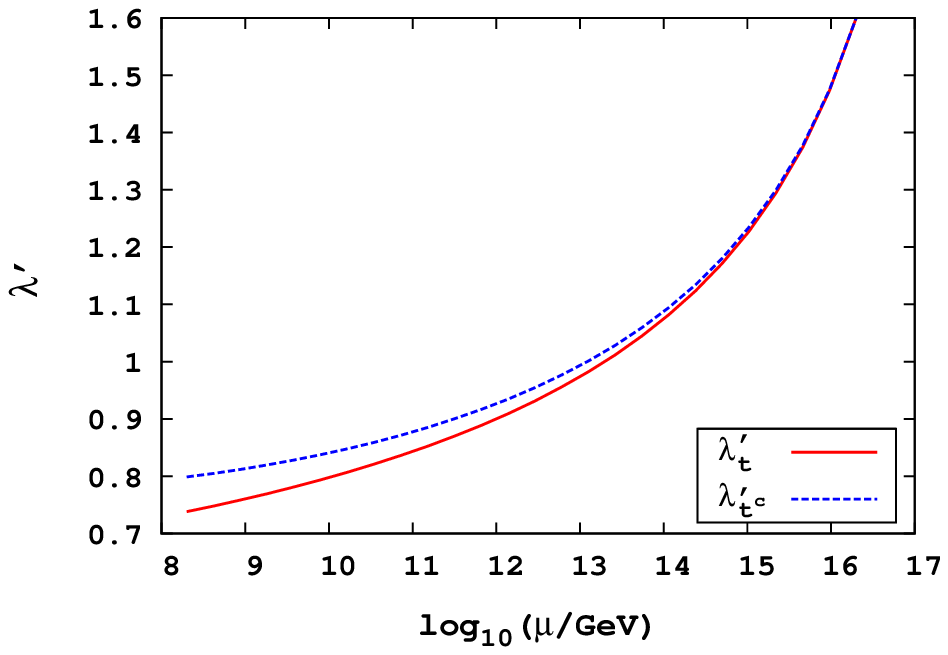}} \caption{
The running of the mixed messenger--matter Yukawa couplings $\lambda'_b$ and $\lambda'_{\tau^c}$ of Eq. (\ref{W5})
of the $5+\overline{5}$ messenger model (left panel). The right panel shows
the evolution of $\lambda'_t$ and $\lambda'_{t^c}$ of
of Eq. (\ref{W10}) of the $10+\overline{10}$ messenger model
from $M_X = 2 \times 10^{16}$ GeV down to
the messenger scale $M_{\rm mes} = 10^8$ GeV. In both cases the unified Yukawa couplings are
taken to be $\lambda'_0=1.6$ and $f_0=0.25$. In the right panel, $\lambda'_{m0} = 0.1 $ has been used.}
\label{RGE}
\end{figure}

In this model, messenger fields belong to $5+\overline{5}$ of $SU(5)$,
with the content $5 = (\overline{d^c}_m + \overline{L}_m)$ and
$\overline{5} = (d^c_m + L_m)$.  Here $d^c_m$ and $L_m$ have the same
quantum numbers as the $d^c$ and $L$ superfields of MSSM. We assume that
these messenger fields have the same $R$--parity as the quarks and leptons
of MSSM.\footnote{While this work was written up
a related work appeared, which discusses messenger mixing with the MSSM Higgs fields,
$W \supset Q_3 u_3^c \overline{L}_m$, along with $H_u-\overline{L}_m$ mixing \cite{ibe}.}
The following $R$--invariant superpotential can now be written, which mixes the MSSM
fields with the messenger fields:
\begin{eqnarray}
W_{5+\overline{5}} &=& f_d\overline{d^c}_md^c_m Z +f_e\overline{L}_mL_m Z +\lambda'_b
Q_3d^c_mH_d+\lambda'_{\tau^c}L_m
e^c_3H_d~.
\label{W5}
\end{eqnarray}
Here we have assumed that the messenger fields couple only with
the third family MSSM fields.  This will be justified based on a
flavor symmetry discussed in Sec. 4, where the lighter family
couplings to the messenger fields are suppressed by powers of
a small parameter $\epsilon$.\footnote{We shall in fact see that
by field redefinitions the general messenger--matter mixing can be brought to
the form of Eq. (\ref{W5}).}  Eq. (\ref{W5}) can arise in
$SU(5)$ as $W = f_0 5_m \overline{5}_m Z + \lambda_0'10_3 \overline{5}_m \overline{5}_H$
with only the $H_d$ component kept from $\overline{5}_H$ (the color
triplet from the $5_H$ and $\overline{5}_H$ acquire GUT scale masses and decouple at $M_X$).  Thus, imposing
$SU(5)$ symmetry, we see that at the GUT scale $M_X \simeq 2 \times 10^{16}$ GeV, there
are only two unified Yukawa couplings $(f_0, ~\lambda'_0)$ that involve
the messenger fields.  The RGE for the Yukawa couplings entering Eq. (\ref{W5}), along with those for
the MSSM Yukawa couplings, are listed in
Appendix A1, valid for the momentum regime $M_{\rm mess} < \mu < M_{X}$.  In Fig. \ref{RGE},
left panel, we plot the evolution of the couplings $\lambda'_b$ and $\lambda'_{\tau^c}$ of Eq. (\ref{W5}) in this momentum regime,
assuming unification of these couplings at the scale $M_X$, and taking $\lambda'_0  = 1.6$
and $f_0=0.25$.

\begin{figure}[h!]
\centering {\includegraphics[width=7.5 cm]{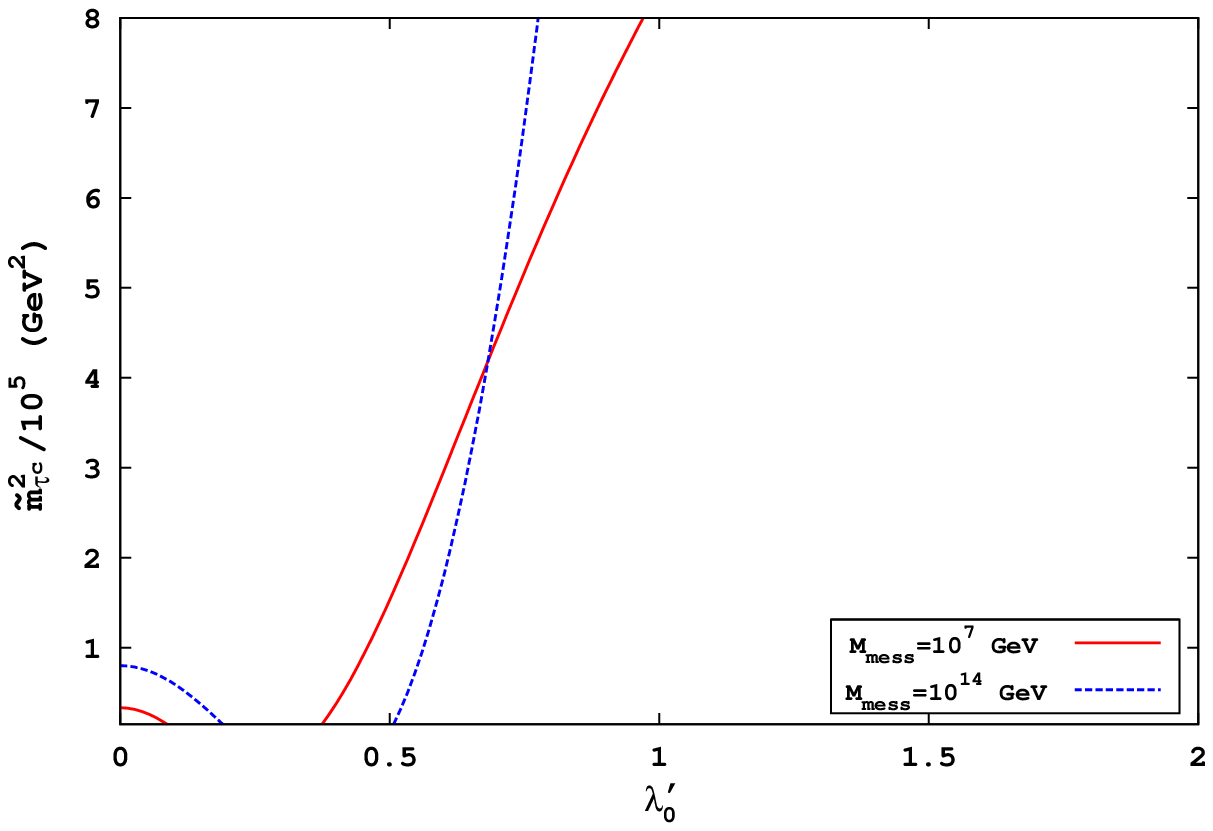}}
{\includegraphics[width=7.5 cm]{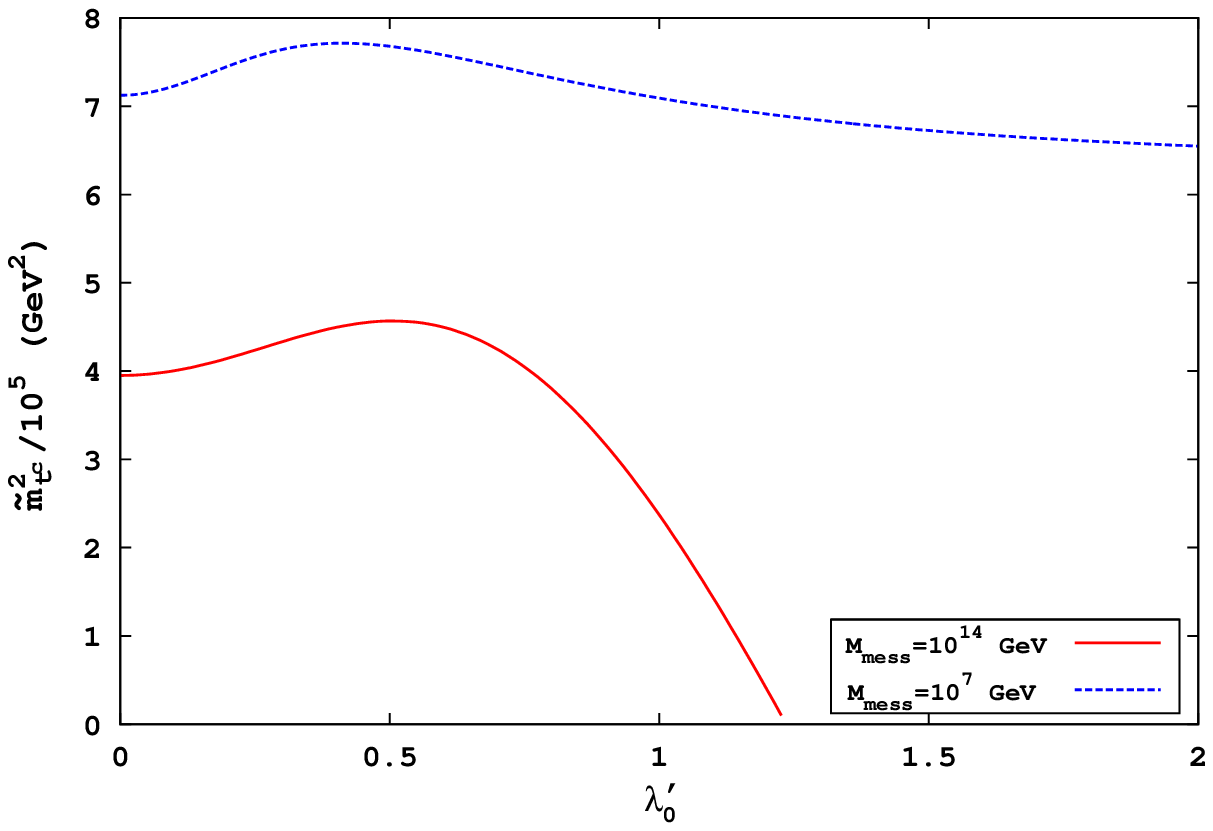}} \caption{
 $\tilde{m}^2_{\tau^c}$ versus $\lambda'_{0}$ at the scale
$M_{\rm{mess}}$ for two different messenger scales $M_{\rm mess} = (10^7,\, 10^{14})$ GeV, in the $5+\overline{5}$ model
(left panel). The right panel shows
$\tilde{m}^2_{t^c}$ versus  $\lambda'_{0}$ at $M_{\rm mess}$ for the same two messenger scales in this model. Here $f_0=0.25$ has been used. }
\label{spec-exclu}
\end{figure}

Without messenger--matter mixing, the scalar masses and the trilinear $A$--terms
at $M_{\rm mess}$ are obtained from Eqs. (\ref{spectrum})-(\ref{AB}).
With such mixings allowed, as in Eq. (\ref{W5}), these relations are modified.
It was shown in Ref. \cite{dine2} that
the mixed messenger--matter couplings would induce negative one-loop
contributions to the supersymmetry-breaking masses. However, these
one-loop contributions have additional $\langle F_Z\rangle/M^2_{\rm mess}$
suppression factors, and can be safely ignored compared to the two--loop
induced terms which do not have such suppression, provided that
$\langle F_Z \rangle/M_{\rm{mess}}^2\leq g_3/4\pi$.  We shall assume that this condition
is met in this paper.  For $M_{\rm mess} > 10^7$ GeV, this condition is
automatically satisfied.  New contributions to the scalar masses and the $A$--terms arise
at the two--loop and one--loop level respectively, proportional to the mixed messenger--matter Yukawa couplings.
These contributions can be obtained from the general expressions
given in Ref. \cite{GR2,Chacko}.   The Yukawa couplings  $\lambda'_b$ and $\lambda'_{\tau^c}$ of Eq. (\ref{W5})
lead to a splitting in the mass of the $\tilde{Q}_3$ squark doublet from those of $\tilde{Q}_{1,2}$, and of the
right--handed stau $\tilde{\tau}^c$ from those of $\tilde{e}^c_{1,2}$.  These shifts, which add to the
universal contributions of Eq. (\ref{spectrum}) at the messenger scale are
(see Appendix B for the derivation):
\begin{eqnarray}
\delta\tilde{m}^2_{Q_3}&=&\frac{\alpha'_{b}\Lambda^2}{8\pi^2}\left(3\alpha'_{b}+\frac{1}{2}\alpha'_{\tau^c}-\frac{8}{3}\alpha_3-\frac{3}{2}\alpha_2-\frac{7}{30}\alpha_1\right),
\label{p43}\\
\delta\tilde{m}^2_{\tau^c}&=&\frac{2\alpha'_{\tau^c}\Lambda^2}{8\pi^2}\left(2\alpha'_{\tau^c}+\frac{3}{2}\alpha'_{b}-\frac{3}{2}\alpha_2-\frac{9}{10}\alpha_1\right),\label{p44}\\
\delta \tilde{m}^2_{H_d}&=&\frac{\delta
\tilde{m}^2_{\tau^c}}{2}+3\delta
\tilde{m}^2_{Q_3}+\frac{3\Lambda^2\alpha'_b\alpha_t}{16\pi^2}
\label{spectrum5}~.
\end{eqnarray}
New contributions to the $A$-terms generated by messenger--matter mixing at the
messenger scale are given by
\begin{eqnarray}
\delta A_t&=&-\frac{1}{4\pi}\alpha'_{b}\Lambda,\label{AB5}\\
\delta A_b&=&-\left(\frac{4\alpha'_{b}+\alpha'_{\tau^c}}{4\pi}\right)\Lambda,\label{p47}\\
\delta
A_{\tau}&=&-\left(\frac{3\alpha'_{b}+3\alpha'_{\tau^c}}{4\pi}\right)\Lambda, \label{AB5pp}
\end{eqnarray}
where $\alpha'_{b}=\frac{\lambda'^{2}_b}{4\pi}$, and
$\alpha'_{\tau^c}=\frac{\lambda'^2_{\tau^c}}{4\pi}$. Here we have followed the definition
${\cal L}_{\rm soft} \supset \lambda_{abc} A_{abc} \tilde{\Phi}_a \tilde{\Phi}_b \tilde{\Phi}_c$ for the trilinear
soft terms, corresponding to the supetpotential $W \supset \lambda_{abc} \Phi_a\Phi_b \Phi_c$.  Since
$\lambda'_b$ and $\lambda'_{\tau^c}$ originate from one unified
coupling $\lambda'_{0}$ as shown in the left panel of
Fig. \ref{RGE}, the scalar mass spectrum at the messenger scale depends on $\lambda'_{0}$,
the messenger scale $M_{\rm{mess}}$, and the effective SUSY breaking
scale $\Lambda$. (There is also a mild dependence on $f_0$ via RGE, we
fix $f_0=0.25$ in our analysis.)  A range of  $\lambda'_{0}$ is excluded since it
leads to negative squared masses for certain sparticles
at the scale $M_{\rm{mess}}$. This range depends on the value of $M_{\rm{mess}}$.
For most part, we consider the range $10^7~{\rm GeV} \leq M_{\rm mess} \leq 10^{14}$ GeV,
the lower value arising from the demand that the negative one--loop contributions to scalar
masses remain small, and the upper value arising by requiring the supergravity contributions
to be small.  We plot the exclusion on $\lambda'_0$ from the positivity of the right--handed stau mass
in Fig. \ref{spec-exclu}, left panel. On the right panel, the right--handed stop mass is plotted, versus
$\lambda'_0$.  Fig. \ref{spec-exclu} shows that the interval $0.2 < \lambda'_{0}< 0.5$ ($0.1 < \lambda'_0 < 0.4$)
is excluded, corresponding to $M_{\rm mess} = 10^{14}$ GeV ($10^7$ GeV),
since that leads to negative
$\tilde{m}^2_{\tau^c}$.  We also
see that both the $\tilde{\tau}^c$ and the $\tilde{t}^c$ can be relative light in this scenario.

Below the scale $M_{\rm{mess}}$, the theory is just the MSSM.
We have solved the one--loop RGEs for the MSSM  with the boundary conditions at
$M_{\rm{mess}}$ given by Eqs. (\ref{spectrum})-(\ref{AB}) and by
Eqs. (\ref{p43})-(\ref{AB5pp}). The soft breaking mass--squared
$m^2_{H_u}$ is driven to negative values at low energy scale, leading
to the breaking of  electroweak symmetry. In order to avoid driving
$\tilde{m}^2_{t^c}$ to negative values at low energy scale,
so that color and electric charge remain unbroken, a region of
$\lambda'_{0}$ is forbidden. For example, the region of
$\lambda'_{0}>1.3$ for $M_{\rm{mess}}=10^{14}$ GeV is forbidden as
shown in the right panel of Fig. \ref{spec-exclu}. This exclusion
arises because of the top quark Yukawa coupling contribution to the
$\tilde{m}^2_{t^c}$, in conjunction with $A_t$ in the RGE,
which becomes large due to the large initial $A_t$ value at $M_{\rm mess}$.

Since all the soft terms at the messenger scale are determined by the
three parameters $\lambda'_{0}$, $\Lambda$ and $M_{\rm{mess}}$ (with $f_0 = 0.25$ fixed
for RGE evolution),
the lightest Higgs mass is also determined by these
three parameters.  As we discussed previously, the maximal mixing
condition $\tilde{A_t}=\sqrt{6}M_s$ (or $X_t=6$) gives the largest value of
the lightest Higgs boson mass. It is not possible to realize this
maximal mixing condition in GMSB without messenger--matter mixing because
$A_t$ vanishes at the scale $M_{\rm{mess}}$ and the induced value at
low energy scale through RGEs is not sufficient. On the other hand,
allowing mixed messenger--matter couplings generates $A_t$ as shown in
Eq. (\ref{AB5}). This leads to an enhancement of the Higgs mass. Choosing
the parameters to lie in the range $4\times
10^4$ $\rm{GeV}$ $<\Lambda <$ $2\times 10^5$ $\rm{GeV}$,  $ 10^7$ $
\rm{GeV}$ $< M_{\rm{mess}} <$ $10^{14}$ $\rm{GeV}$ and
$0<\lambda'_{0} <2$, we report the numerical values for the lightest Higgs boson
mass $m_h$ in Table \ref{mhvalue} for different choices of these parameters.
In this Table, we have excluded values of $\lambda'_{0}$
that give negative values for $\tilde{m}^2_{\tau^c}$ and
$\tilde{m}^2_{t^c}$. We have used Eq. (\ref{higgsmass}) to compute $m_h$,
and since SuSpect gives $m_h$ values systematically higher by 2 GeV, the
quoted upper limit of $m_h = 114$ GeV for the case of $\lambda'_0=0$ actually
should be interpreted as $m_h = 116$ GeV.   This value increases by
about 5 GeV to $121$ GeV in the case of large $\lambda'_0$.
This limit is 118 GeV when the stops have masses $< 1.5$ TeV, as
indicated in Table \ref{mhvalue}.  While the increase in $m_h$ is significant
in this model with messenger--matter mixing, here maximal stop mixing is
not realized, primarily due to the positivity conditions on $\tilde{m}_{t^c}^2$.
Note that there is no contribution to  $\tilde{m}_{t^c}^2$ from the mixed
Yukawa coupling in this model, which implies that this parameter turns negative
quickly below $M_{\rm mess}$ if $\lambda_0'$ is large. This situation improves, enabling larger values for $m_h$
when the messenger fields belong to $10+\overline{10}$, as discussed in the next subsection.

\begin{table}[h!]
 \center
\begin{tabular}{|c|c|c|c|c|c|} \hline
$\lambda'_{0}$ &$m_h (\rm{GeV})$
&$\Lambda(10^5\rm{GeV})$&$M(10^{13}\,\rm{GeV})$&$\tilde{m}_{t_1}(\rm{GeV})$&
$\tilde{m}_{t_2}(\rm{GeV})$\\\hline
 0&114&2&1.78&1249&1695\\\hline
 0.8&116&2&10&1212&1583\\\hline
 1.2&119&2&10&384&2613\\\hline
\end{tabular}
\caption{The lightest Higgs boson mass $m_h$ in the $5+\overline{5}$ model as functions
of the GMSB input parameters, $\Lambda$, $\lambda_0'$ and $M_{\rm{mess}}$ for
$\tan\beta=10$. Here we have fixed $f_0 = 0.25$. }\label{mhvalue}
\end{table}

\subsection{Higgs mass bound in the $10+\overline{10}$ messenger model}

Here we consider messenger fields belonging to
$10+\overline{10}$ of $SU(5)$. These fields decompose in terms of MSSM--like
fields as:
\begin{eqnarray}
10+\overline{10}=(Q_m+\overline{Q}_m)+(u^c_m+\overline{u^c}_m)+(e^c_m+\overline{e^c}_m).
\end{eqnarray}
As before, we assume that  the messenger fields only couple with the third
generation of MSSM fields, and that they have the same $R$--parity as the MSSM quarks and leptons.
In this case the following superpotential couplings can be written.
 \begin{eqnarray}
W_{10+\overline{10}}&=&\lambda'_{t^c}Q_3u^c_mH_u + \lambda'_{t}Q_mu^c_3H_u+\lambda'_m Q_m u^c_mH_u\nonumber\\
&+&f_{e^c}\overline{e^c}_me^c_mZ+f_{u^c}\overline{u^c}_mu^c_mZ+f_{Q}\overline{Q}_mQ_mZ.
\label{W10}
\end{eqnarray}
Although the couplings $Q_m\, d_3^c\, H_d + L_3 \,e^c_m\, H_d$ are allowed by gauge symmetry, we
have not included them in the above superpotential because these terms will be
suppressed by a small parameter $\epsilon$ when this model is embedded in a flavor $U(1)$ symmetric
framework, as we shall see in the next section. The couplings of Eq. (\ref{W10}) can arise in $SU(5)$ theory from
$W \supset \lambda'_0 10_3 10_m 5_H + \lambda'_{m0} 10_m 10_m 5_H + f_0 10_m \overline{10}_m Z$, with
only the $H_u$ component of $5_H$, and not the color triplet component, kept below $M_X$. Thus we see that the Yukawa couplings
$\lambda'_{t^c}$ and $\lambda'_{t}$ are equal to the unified
coupling $\lambda'_{0}$ at the GUT scale. Similarly, the three Yukawa couplings $f_{e^c}$,
$f_{Q}$ and $f_{u^c}$ are equal to a single coupling $f_0$ at the GUT scale.
In other words, the six Yukawa couplings appearing in the
superpotential of Eq. (\ref{W10}) are reduced to three:
$\lambda'_{0}$, $f_0$ and $\lambda'_{m0}$ at the GUT scale. We shall use these unification conditions
and derive the couplings of Eq. (\ref{W10}) by using the RGE listed in Appendix A2.
The evolution of $\lambda'_t$ and $\lambda'_{t^c}$ below
$M_X$ is shown in the right
panel of Fig. \ref{RGE} with $f_0 = 0.25$ and $\lambda'_{m0} = 0.1$ fixed.

The Yukawa couplings $\lambda'_{t^c}$, $\lambda'_{t}$ and
$\lambda'_{m}$ generate 2-loop (1-loop) scalar masses ($A$-terms) at
the scale $M_{\rm{mess}}$, as derived in Appendix B2.  As a result, the
universal scalar masses given by Eqs. (\ref{spectrum}) and (\ref{AB})
(with $N_{\rm{mess}}=3$ corresponding to $10+\overline{10}$ messenger fields) would receive additional contributions at the
scale $M_{\rm{mess}}$:
\begin{eqnarray}
\delta\tilde{m}^2_{Q_3}&=&\frac{\Lambda^2}{8\pi^2}\left[\alpha'_{t^c}\left(3\alpha'_{t^c}+\frac{3}{2}\alpha'_{t}+\frac{5}{2}\alpha'_m-\frac{8}{3}\alpha_3-\frac{3}{2}\alpha_2-\frac{13}{30}\alpha_1\right)\right.\nonumber\label{p418}\\
&-&\left.\alpha_t\left(\frac{5}{2}\alpha'_{t}+\frac{3}{2}\alpha'_m\right)\right],
\label{spectrum10p}\\
\delta\tilde{m}^2_{t^c}&=&\frac{2\Lambda^2}{8\pi^2}\left[\alpha'_{t}\left(3\alpha'_{t}+\frac{3}{2}\alpha'_{t^c}+2\alpha'_m-\frac{8}{3}\alpha_3-\frac{3}{2}\alpha_2-\frac{13}{30}\alpha_1\right)\right.\nonumber\label{p419}\\
&-&\left.\alpha_t\left(2\alpha'_{t^c}+\frac{3}{2}\alpha'_m
\right)\right],\\
\delta\tilde{m}^2_{H_u}
&=&\frac{3\Lambda^2}{8\pi^2}\left[\alpha'_{t^c} \left(3\alpha'_{t^c}
+\frac{3}{2}\alpha'_{t}+\frac{5}{2}\alpha'_m -\frac{8}{3}\alpha_3
-\frac{3}{2}\alpha_2-\frac{13}{30}\alpha_1\right)\right.\nonumber\\
&+&\left.\alpha'_{t}\left(3\alpha'_{t}+\frac{3}{2}\alpha'_{t^c}+2\alpha'_m-\frac{8}{3}\alpha_3-\frac{3}{2}\alpha_2-\frac{13}{30}\alpha_1\right)\right.\nonumber\\
&+&\left.\alpha'_m\left(3\alpha'_m+2\alpha'_{t}+\frac{5}{2}\alpha'_{t^c}-\frac{8}{3}\alpha_3-\frac{3}{2}\alpha_2-\frac{13}{30}\alpha_1
\right)\right],
\end{eqnarray}
\begin{eqnarray}
\delta A_t&=&-\left[\frac{5\alpha'_{t}+4\alpha'_{t^c}+3\alpha'_m}{4\pi}\right]\Lambda,\label{AB10}\\
\delta A_b&=&-\frac{\alpha'_{t^c}}{4\pi}\Lambda\label{AB10p},
\end{eqnarray}
where $\alpha'_{t^c}=\frac{\lambda'^{2}_{t^c}}{4\pi}$,
$\alpha'_{t}=\frac{\lambda'^{2}_{t}}{4\pi}$, and
$\alpha'_{m}=\frac{\lambda'^{2}_{m}}{4\pi}$. An interesting feature
of the $10+\overline{10}$ model is that unlike the $5+\overline{5}$ model, here along with $A_t$, the $\tilde{m}^2_{t^c}$
also receives new contributions which can be positive.  As a result, sufficiently large $A_t$ can be generated without
turning $\tilde{m}_{t^c}^2$ negative, and
the maximal mixing condition $X_t=6$ can be realized, leading to an increased upper limit
on $m_h$, as large as $(125-126)$ GeV.

In order to find the upper limit on $m_h$ and the SUSY mass spectrum,
we solve the MSSM RGE numerically from the messenger scale to the
low scale with the boundary conditions given in Eqs. (\ref{spectrum10p})-(\ref{AB10p})
and in Eqs. (\ref{spectrum})-(\ref{AB}).
These masses would depend on  four parameters:
$\Lambda$, $M_{\rm{mess}}$, $\lambda'_{0}$ and $\lambda'_{m0}$. (The value of $f_0$ is
also relevant for RGE evolution, we fix $f_0=0.25$ in our analysis. $m_h$ is not
very sensitive to the choice of $f_0$.)
In Table \ref{p4tab2} we report the values of $m_h$ for different
values of $\Lambda$, $M_{\rm{mess}}$ and $\lambda'_{0}$ with
a fixed value of $\lambda'_{m0}=0$.  In Table \ref{p4tab3} we report
the same, but now with $\lambda'_{m0}=1.2$ fixed.  In both cases
$m_h = 125$ GeV can be obtained (once 2 GeV is added to the numbers
quoted in these tables), with all SUSY particles below 1.5 TeV.
For example, in the case of $\lambda'_{m0}=0$ (Table \ref{p4tab2}),
without messenger--matter mixing, obtaining $m_h = 119$ GeV would require
one of the stops to be heavier than 3 TeV, while with such mixings,
$m_h = 125$ GeV is realized with both stops below 1.5 TeV.

\begin{table}[h!]
\center
\begin{tabular}{|c|c|c|c|c|c|c|} \hline
$\lambda'_0$ &$m_h (\rm{GeV})$
&$\Lambda(10^5\rm{GeV})$&$M_{\rm{mess}}(\rm{GeV})$&$\tilde{m}_{t_1}(\rm{GeV})$&
$\tilde{m}_{t_2}(\rm{GeV})$&$A_t/M_s$\\\hline
 0&117&1.6&$3 \times10^{13}$&2656&3284&$-0.86$\\\hline
 0.4&118&1.36& $10^8$ &1795&2396&$-1.27$\\\hline
 0.8&122 &0.912& $10^{13}$ &1553&2143&$-1.95$\\\hline
1.1&123&0.784& $2\times10^{11}$ &735&1429&$-2.0$ \\\hline
2&123&0.784& $10^8$ &743&1426&$-2.26$\\\hline
\end{tabular}
\caption{The lightest Higgs boson mass $m_h$, along with the stop masses,
and the stop mixing parameter $A_t/m_s$ for different values of the GMSB input parameters $\Lambda$,
$\lambda'_0$ and $M_{\rm{mess}}$ in the $10+\overline{10}$ model.  Here we have fixed $\lambda'_{m0}=0$,
$f_0=0.25$, and set
$\tan\beta=10$.} \label{p4tab2}\addvspace{0.5cm}
\end{table}

\begin{table}[h!]
\center
\begin{tabular}{|c|c|c|c|c|c|c|} \hline
$\lambda'_0$ &$m_h
(\rm{GeV})$&$\Lambda(10^5\rm{GeV})$&$M_{\rm{mess}}(\rm{GeV})$
&$\tilde{m}_{t_1}(\rm{GeV})$&
$\tilde{m}_{t_2}(\rm{GeV})$&$A_t/M_s$\\\hline
 0&121&0.97&$2 \times10^{13}$ &928&1636&$-1.8$\\\hline
 0.4&123&0.91&$3 \times10^{13}$&656&1612&$-2.3$\\\hline
 0.6&123&0.848& $10^{12}$ &673&1512&$-2.3$\\\hline
0.8&123&0.784& $10^{11}$ &682&1509&$-2.3$\\\hline
2&123&0.784& $10^8$ &753&1425&$-2.2$\\\hline
\end{tabular}
\caption{Same as in Table \ref{p4tab2}, but now with $\lambda'_{m0}=1.2$ fixed.} \label{p4tab3}\addvspace{0.5cm}
\end{table}

\begin{figure}
\centering {\includegraphics[width=7.5 cm]{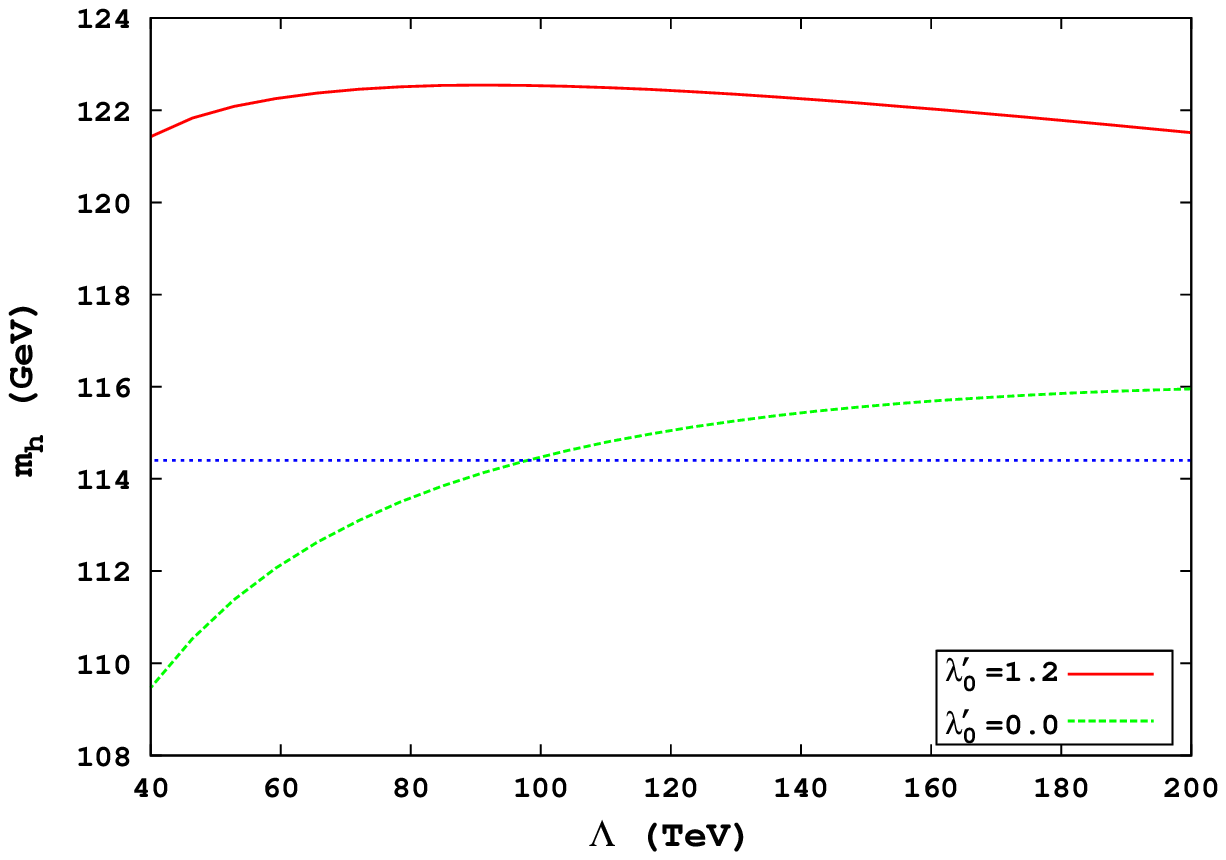}}
{\includegraphics[width=7.5 cm]{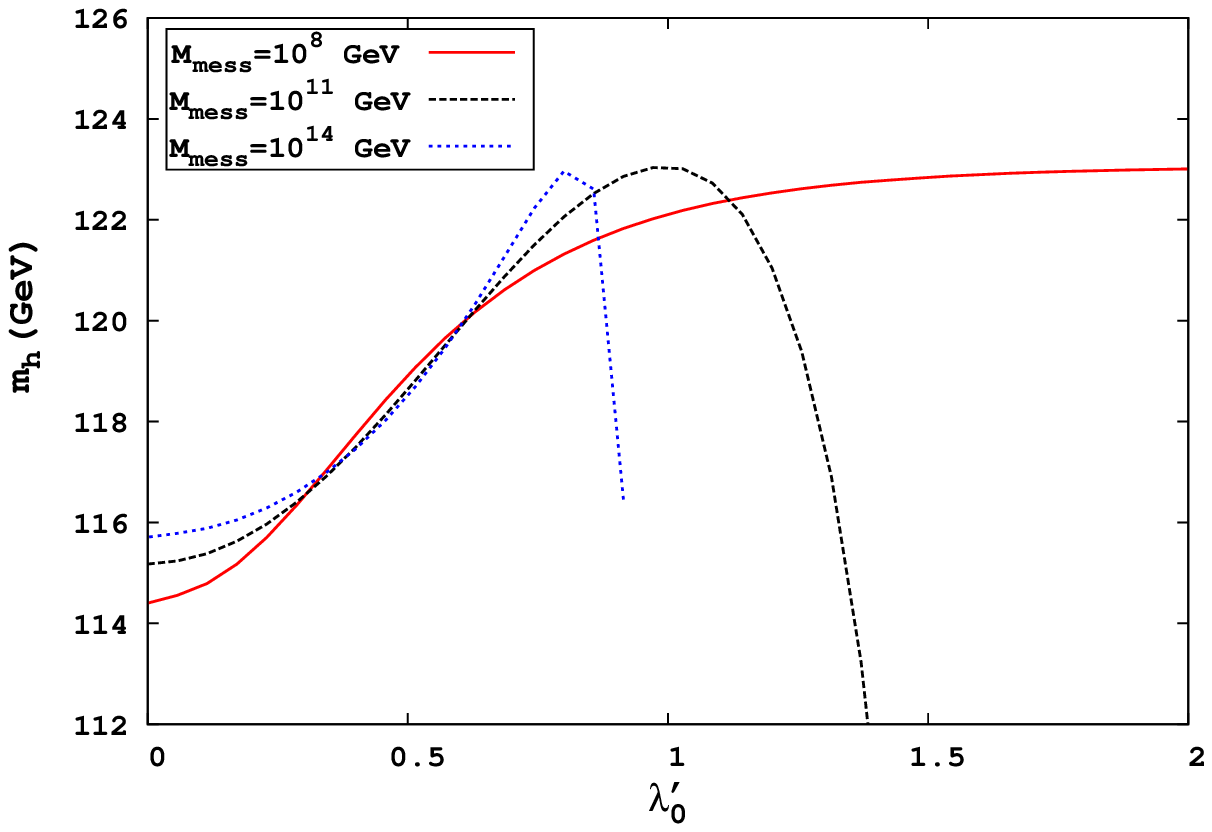}} \caption{ $m_{h}$ versus $\Lambda$ for $\lambda'_{0}=0$
and $\lambda'_{0}=1.2$ (left panel). The horizontal line indicates the LEP
lower limit $m_h > 114.4$ GeV. The right panel shows $m_{h}$ versus
$\lambda'_{0}$ for different messenger scales and with $\Lambda=10^5$
GeV fixed.}
\label{p4fig3}
\end{figure}

In Fig. \ref{p4fig3}, left panel, we plot
the  Higgs mass as a function of $\Lambda$ for two values of the unified Yukawa
coupling $\lambda'_0 = (0, 1.2)$, where $\lambda'_0=0$ corresponds to minimal GMSB without
messenger--matter mixing.  We see that the Higgs mass is raised by
10 GeV in the case of $\lambda'_0=1.2$ compared to the case of $\lambda'_0=0$
for low values of $\Lambda = 4 \times 10^4$ GeV.  This increase is
about 6 GeV for larger $\Lambda$.  Note that smaller values of $\Lambda$
leads to lighter SUSY particles, with the stop mass around $500-600$ GeV,
which might be accessible to early run of LHC. In the right panel of
Fig.  \ref{p4fig3} we have plotted $m_h$ versus $\lambda'_0$ for various
values of $M_{\rm mess}$, and for $\Lambda = 10^5$ GeV fixed.
There is a non-trivial constraint on $\lambda'_0$ when $M_{\rm mess} > 10^{11}$ GeV,
owing to the stop squared mass turning negative at low energies.
Note that $m_h \simeq 125$ GeV is realized in this model, along with sub--TeV superparticles,
even for low messenger scale,  $M_{\rm{mess}}\leq3\times 10^8$ $\rm{GeV}$, preferred by cosmology.

We present three different spectra  for the superparticle masses in Table \ref{p4tab4}, two corresponding
to the $10+\overline{10}$ model, and one for the $5+\overline{5}$ model of the previous subsection.
In this Table, the masses quoted in the last two columns correspond to $\tan\beta=6.1$ (for the
$10+\overline{10}$ model) and $\tan\beta=15.6$ (for the $5+\overline{5}$ model).  These values
are derived by assuming the vanishing of the $B$--term at $M_{\rm mess}$, as in Eq. (\ref{AB}).
The third column of Table \ref{p4tab4} lists the sparticle spectrum for an arbitrary value
of $\tan\beta = 10$.  The spectrum in the fourth column assumes a low messenger scale of $M_{\rm mess} =
4 \times 10^5$ GeV.  The negative one--loop contributions to the scalar masses are $< 5\%$ of the
positive two--loop contributions arising from messenger--matter mixing for this value of $M_{\rm mess}$.
For larger values of $M_{\rm mess}$, as in the third and fifth columns of Table \ref{p4tab4}, these
one--loop contributions are even smaller.  The mass values of Table \ref{p4tab4} show that light SUSY
spectrum is possible in GMSB along with a Higgs boson mass around 125 GeV, if messenger--matter mixing is allowed.
We shall use the mass values of Table \ref{p4tab4} in deriving flavor violation constraints on the model, which
is addressed in the next section.

\begin{table}[h!]
\center \small\addtolength{\tabcolsep}{4pt}
\scalebox{0.9}{%
\begin{tabular}{|c|c|c|c|c|} \hline
 Particle &                                                 &$10+\overline{10}$   &  $10+\overline{10}$       &$5+\overline{5}$   \\\hline\hline
 Inputs      &  $M_{\rm{mess}}$                              &    $10^8 $          &$4 \times 10^5$            &   $10^8 $   \\
             &  $N_{\rm{mess}}$                              &   3                 &3                          &    1         \\
             &  $\Lambda(10^5\,\rm{GeV})$                        &   $ 0.45$        & $ 0.3$          &    $1.5$  \\
             &  $\tan\beta$                             &   10                &6.1                        &15.6                    \\
               &  $f_0$                             &  0.25              &0.25                      &0.25                  \\
             &   $\lambda_{0}$                          &  1.3                &1.2                        &    1.2              \\\hline\hline
Higgs:      & $m_h$                                     & 122                 &118                      &  114.5  \\
            & $m_H^0$                                   & 858                 &592                        & 1690  \\
            & $m_A$                                     & 858                 &591                        &  1690 \\
            & $m_{H^{\pm}}$                             & 862                 &597                        &  1689\\\hline\hline
 Gluino:   & $\tilde{m}_{{g}}$                            & 980               &667                        & 1041\\\hline\hline
 Neutralinos:& $m_{\chi_1}$                             & 186                 &124                        & 208\\
             & $m_{\chi_2}$                             & 346                 &225                        & 408\\
             & $m_{\chi_3}$                             & 800                  &557                        & 781\\
             & $m_{\chi_4}$                             & 807                 &569                        & 790\\\hline\hline
Charginos:    & $\chi^+_1$                              & 347                  &227                        & 409\\
             & $\chi^+_2$                               & 807                  &569                        & 790\\\hline\hline
Squarks:      & $\tilde{m}_{{u}_L,{c}_L}$           & 972                  &657                        &1480\\
              & $\tilde{m}_{{u}_R,{c}_R}$           & 929                  &632                        & 1377\\
              & $\tilde{m}_{{d}_L,{s}_L}$           & 971                  &657                        &1480\\
              & $\tilde{m}_{{d}_R,{s}_R}$           & 922                  &630                        & 1365\\
              & $\tilde{m}_{{b}_L}$                       & 800                  &555                        & 1315\\
              & $\tilde{m}_{{b}_R}$                       & 919                  &629                        & 1294\\
              & $\tilde{m}_{{t}_L}$                       & 853                  &621                        & 1315\\
              & $\tilde{m}_{{t}_R}$                       & 412                  &270                        & 1123\\\hline\hline
Sleptons:     & $\tilde{m}_{{e}_L,{\mu}_L}$        & 323                           &200                       &596\\
              &$\tilde{m}_{{\nu_e}_L,{\nu_{\mu}}_L}$& 323                  &200                       &596\\
              & $\tilde{m}_{{e}_R,{\mu}_R}$         & 152                  &92                        &290\\
              & $\tilde{m}_{{\tau}_L}$                    & 322                  &197                        &539\\
              & $\tilde{m}_{{\tau}_R}$                    & 151                  &92                        & 1543\\\hline \hline
\end{tabular}}
\caption{The SUSY spectrum corresponding to $10+\overline{10}$ model and
$5+\overline{5}$ model for three choices of input parameters. All masses are in GeV.
The values of $\tan\beta$ in the last two columns are derived from the condition that $B=0$ at $M_{\rm mess}$.
2 GeV should be added to $m_h$ quoted here to be consistent with results obtained from SuSpect.}\label{p4tab4}
\end{table}

\section{Flavor violation induced by messenger--matter mixing}

The main motivation for gauge mediation of SUSY breaking is
that it naturally solves the SUSY flavor problem.  This is
possible because of the universality in the scalar masses
induced by gauge mediation.  This universality is however
violated by messenger--matter mixing, as seen from Eq. (\ref{spectrum5})
in the $5+\overline{5}$ model, and from Eq. (\ref{spectrum10p}) in the
$10+\overline{10}$ model.  In this section we show that flavor violation induced
by such non--universal contributions to the soft scalar masses can be all
within experimental limits, if we embed these models in a framework with
a $U(1)$ flavor symmetry.  This $U(1)$ symmetry also addresses the hierarchies
in the fermion masses and mixings \cite{fn}.  When embedded in $SU(5)$ unified theory,
this framework would lead to a lopsided structure for the down quark and
charged lepton mass matrices \cite{lopsided}, which explains naturally why the
quark mixing angles are small, while the leptonic mixing angles are large.
Such matrices also explain other features of the fermion mass spectrum, such
as why the charge $2/3$ quark mass ratios exhibit a stronger hierarchy
compared to the charge $-1/3$ quark mass ratios or the charged lepton mass ratios.
We shall see that while dangerous flavor violation is suppressed by this flavor
symmetry, small amount of flavor violation is present in these models, which can
have testable consequences.

The flavor $U(1)$ symmetry serves another important purpose.  It forbids bare masses for
the messenger fields, a requirement for successful gauge mediation.  GMSB models
usually assume these bare masses are zero, here there is a symmetry based explanation
for them to vanish.

In our construction, owing to the $U(1)$ flavor symmetry, renormalizable Yukawa couplings are allowed
only for the third family fermions.   The vacuum expectation value
of a SM singlet field $S$, which  breaks this $U(1)$ at a scale slightly below  $M_*$,
identified as the Planck scale or the string scale, generates masses for the first two families
via non-renormalizable operators which are suppressed by powers of a small parameter
$\epsilon \equiv \langle S \rangle/M_*$.  The power suppression arises because
of the flavor--dependent $U(1)$ charges of the fermions.  In such a framework, all
fundamental Yukawa couplings can be of order one and still the hierarchy in the fermion masses
and mixings can be explained \cite{babuTASI}.  This $U(1)$ can be
naturally identified as the anomalous $U(1)$ symmetry of string theory \cite{gs}.

We now turn to the embedding of the the $5+\overline{5}$ messenger model and the
$10+\overline{10}$ messenger model of the previous section into a unified $SU(5)$ framework
along with a flavor $U(1)$ symmetry and discuss flavor violation
mediated by SUSY particles in these models.

\subsection{Flavor violation in the $5+\overline{5}$ messenger model}

Although we do not construct complete $SU(5)$ models, the assignment of $U(1)$
charges for the fields will be compatible with $SU(5)$ symmetry.  So we can use the
notation of SUSY $SU(5)$.  The three families of quarks and leptons belong to
$\overline{5}_i+10_i$ under $SU(5)$, with $i=1-3$.  Here $10_i \subset \{Q_i,\,u_i^c,\,e_i^c\}$
and $\overline{5}_i \subset \{d^c_i,\,L_i\}$.  The Higgs doublets ($H_u,\,H_d$) of MSSM are
contained in $5_H$ and $\overline{5}_H$ of $SU(5)$.  It should be understood that from
these Higgs fields, the color triplet components have been removed for our discussions
which relate to momentum scales below $M_X$.  The messenger fields
are denoted as $5_m+\overline{5}_m$, and are assumed to have the same $R$--parity as quarks
and leptons.  The flavor $U(1)$ charges of these fields are listed in Table \ref{charge5}.  The
charge assignment for the MSSM fields is the same as the one given in Ref. \cite{enkhbat}, but
here we extend it to include the messenger fields.

\begin{table}[h!]
\center
\begin{tabular}{|c|c|c|c|c|c|c|c|c|c|c|} \hline
 Particle &$10_1$ &$10_2$&$10_3$& $\overline{5}_1$ &$\overline{5}_2$, $\overline{5}_3$& $5_H$, $\overline{5}_H$&$S$&$5_m$& $\overline{5}_m$& $Z$    \\\hline
 $U(1)$ & 4     & 2    &  0   &      $p+1$           &             $p$               &           0           &$-1$ & $-\alpha$  &         0       &  $\alpha$      \\\hline
\end{tabular}
\caption{The $U(1)$ charges of the MSSM fields, the messenger fields, and the singlets $Z$ and
$S$ in the $5+\overline{5}$ messenger model in the $SU(5)$ notation. $p$ here is an integer which
can take values $p=(0,1,2)$ corresponding to (large, medium, small) $\tan\beta$.}\label{charge5}\addvspace{0.5cm}
\end{table}

In Table \ref{charge5} the parameter $p$
is an integer which can take values 0, 1 or 2, corresponding to large, medium or small $\tan\beta$
values. Although the value $p=0$ can explain the fermion mass hierarchy,
we will see that this choice is disfavored from FCNC constraints, while $p=1,2$ are both acceptable.
The field $S$ acquires a
VEV just below $M_*$ without breaking SUSY, while the field $Z$ acquires a VEV along its scalar
component $\langle Z \rangle \sim M_{\rm mess}$, which is much smaller than $\langle S \rangle$.  The field
$Z$ also acquires an $F$--component, which breaks supersymmetry.  If the $U(1)$ symmetry is identified as
the anomalous $U(1)$ of string theory, even without writing any superpotential, $\langle S \rangle \neq 0$ can
develop by the shift in fields required to set the gravity--induced Fayet--Iliopoulos $D$--term for the $U(1)$ to zero,
so that SUSY remains unbroken \cite{dsw}.  In such schemes, typically one finds $\epsilon \equiv  \langle S \rangle/M_*  \sim 0.2$,
which provides a small expansion parameter to explain the fermion mass hierarchy.
The charge $\alpha$ in Table \ref{charge5} is not specified for now, but it should be positive, and if it is an
integer, $\alpha> p+1$ should be satisfied.  These conditions are needed to guarantee that bare masses for the messenger
fields are forbidden, and that the $5_m$ messenger field does not acquire a mass by pairing
with $\overline{5}_i$ fields through superpotential couplings such as $\overline{5}_15_m S^n$
for some positive integer $n$.  (Successful gauge mediation requires that the masses of the
messenger fields arise from the coupling $\overline{5}_m 5_m Z$, with $Z$ acquiring VEVs along the scalar
and $F$--components.) If the charge $\alpha$ is undetermined, that could lead to an additional global $U(1)$
symmetry which can result in an unwanted Goldstone boson. Since $Z$ carries a charge $\alpha$, and
since it couples to the secluded
sector where supersymmetry breaks dynamically, $\alpha$ may get determined
from such couplings.  We also note that for a rational value of $\alpha$,  $\alpha = a/b$ with $a,\,b$ being positive integers,
the superpotential coupling $S^a Z^b/M_*^{b-1}$ is allowed, which can fix $\alpha$ without upsetting the success of gauge
mediation.  For example, the superpotential coupling $S^4 Z^5/M_*^6$ would fix $\alpha=4/5$, which should be
harmless as far as the conditions $\langle F_Z\rangle  \neq 0$, $\langle Z\rangle  \neq 0$ are concerned.

The superpotential of the model consistent with the flavor $U(1)$ symmetry of Table \ref{charge5}
(in the notation of MSSM fields) is
\begin{eqnarray}
W &=& y^u_{ij}\, \epsilon^{n^u_{ij}}\, u^c_i Q_j H_u +  y^d_{ij}\, \epsilon^{n^d_{ij}}\, d^c_i Q_j H_d +
 y^e_{ij}\, \epsilon^{n^e_{ij}}\, e^c_i L_j H_d \nonumber \\
&+& f_d\overline{d^c}_m d^c_m Z +f_e\overline{L}_mL_m Z +\lambda'_b Q_3d^c_mH_d+\lambda'_{\tau^c} L_me_3^cH_d~.
\label{lopsided5}
\end{eqnarray}
Here $y^{u,d,e}_{ij}$ are order one Yukawa couplings.  The powers of $\epsilon$ appear in Eq. (\ref{lopsided5})
from $(\langle S \rangle/M_*)^{n_{ij}}$ factors, needed to preserve the $U(1)$ symmetry.  Here $n^u_{ij} =
Q(u^c_i)+Q(Q_j)$, $n^d_{ij} = Q(d^c_i)+Q(Q_j)$, and $n^e_{ij} = Q(e^c_i)+Q(L_j)$, where $Q(f)$ refers to the
$U(1)$ charge of the field $f$.  Thus $n^d_{12} =  p+3 = n^e_{21}$, etc.

The second line of Eq. (\ref{lopsided5}) represents messenger--matter mixing allowed by the $U(1)$ symmetry.
One can choose a basis where such mixings involve only the third family fermions $Q_3$ and $e_3^c$.
The coupling $\overline{L}_m (f_e'L_m + f_3\epsilon^p L_3 + f_2\epsilon^p L_2 +
f_1\epsilon^{p+1} L_1)\,Z$ has been redefined simply as $f_e \overline{L}_m L_m Z$ by rotating the
($L_i$, $L_m$) fields.  In the $L_\alpha e_j^c H_d$ couplings, terms with $\alpha,\, j = 1-3$ are part
of the first line of Eq. (\ref{lopsided5}), while in the terms $L_m e^c_j H_d$, a redefinition of
$e_j^c$ fields can be made so that a single term $L_m e_3^c H_d$, the last term of Eq. (\ref{lopsided5}),
is necessary.  Similar arguments apply for the $Q$ fields, so that only $Q_3$ has mixed couplings with $d_m^c$.
If $SU(5)$ boundary conditions are applied to the messenger Yukawa couplings, we would have $\lambda'_{\tau^c}
= \lambda'_b$ at $M_X$.  In this section we shall allow for the possibility that these couplings are not
unified, and define a prameter
\begin{eqnarray}
r = \frac{\lambda'_{\tau^c} (M_X)}{ \lambda'_b(M_X)}~
\end{eqnarray}
so that $r=1$ corresponds to $SU(5)$ unification condition, while $r=0$ would imply that $\lambda_{\tau^c}'=0$
at $M_X$ and below. The latter choice will turn out to be useful to satisfy FCNC constraints.  Note that even
when $r=0$, the increase in $m_h$ found in Sec. 3 will hold, since the initial condition for $A_t$ is determined
by $\lambda'_b$ (see Eq. (\ref{AB5})).

The first line
of Eq. (\ref{lopsided5}) provides an explanation for the hierarchy in the masses and mixings of quarks and leptons.
The mass matrices for the up--quarks, down--quarks and charged leptons arising from Eq. (\ref{lopsided5}) have
the form:
\begin{eqnarray}
M^u = Y^uv_u = \left(
\begin{array}{ccc}
y^u_{11}\epsilon^8 & y^u_{12}\epsilon^6 & y^u_{13}\epsilon^4 \\
y^u_{21}\epsilon^6  & y^u_{22}\epsilon^4 & y^u_{23}\epsilon^2 \\
y^u_{31}\epsilon^4  & y^u_{32}\epsilon^2 & y^u_{33}
\end{array} \right) v_u~, \label{Mu}\\
M^d = Y^d v_d =\epsilon^p \left(
\begin{array}{ccc}
y^d_{11}\epsilon^5 & y^d_{12}\epsilon^3 & y^d_{13}\epsilon \\
y^d_{21}\epsilon^4  & y^d_{22}\epsilon^2 & y^d_{23} \\
 y^d_{31}\epsilon^4  & y^d_{32}\epsilon^2 & y^d_{33}
\end{array} \right) v_d ~, \label{Md}\\
M^e = Y^e v_d =\epsilon^p \left(
\begin{array}{ccc}
y^e_{11}\epsilon^5 & y^e_{12}\epsilon^4& y^e_{13}\epsilon^4 \\
y^e_{21}\epsilon^3  & y^e_{22}\epsilon^2 & y^e_{23}\epsilon^2 \\
 y^e_{31}\epsilon & y^e_{32}& y^e_{33}
\end{array} \right) v_d~. \label{Me}
\end{eqnarray}
These matrices have been written down with the left--handed
anti-fermion fields multiplying  on the left and the left--handed fermion fields
multiplying on the right.  With all the $y^{u,d,e}_{ij}$ factors
being order one, we see that these matrices lead to the mass hierarchy
$m_u: m_c: m_t \sim \epsilon^8: \epsilon^4: 1$, $m_d:m_s:m_b \sim
\epsilon^5: \epsilon^2: 1$, and $m_e: m_\mu:m_\tau \sim \epsilon^5: \epsilon^2:1$,
in nice agreement with observations \cite{lopsided,babuTASI}, with the choice $\epsilon \simeq 0.2$.
This pattern explains why the up--type quarks exhibit stronger hierarchy
compared to the down--type quarks, which have a similar hierarchy structure
as the charged leptons. We also see from the (3,3) entries of $M^u$ and $M^d$
that  $\tan\beta \sim \epsilon^p \, ( m_t/m_b)$,
which suggests the values of $p=(0,1,2)$ for (large, medium, small) $\tan\beta$.
Note that the rotations done in obtaining Eq. (\ref{lopsided5}) so that only the third family
couples to messenger fields do not upset the
hierarchy factors of Eqs. (\ref{Mu})-(\ref{Me}).  The mixed Yukawa couplings of Eq. (\ref{lopsided5})
also do not affect these mass matrices, since these contributions are suppressed by
$\lambda' v_d/M_{\rm mess}$.

One can diagonalize the mass matrices of Eqs. (\ref{Mu})-(\ref{Me}) via bi-unitary
transformations defined as $(U_R^{u,d,e})\, M^{u,d,e}\, (U_L^{u,d,e})^ \dagger = M^{u,d,e}_{\rm diag}$.
Then the left--handed rotation matrices $U_L^{u,d,e}$ and the right--handed
rotation matrices $U_R^{u,d,e}$ would be of the form
\begin{eqnarray}
 U^e_L\sim U^d_R &\sim& \left(
\begin{array}{ccc}
1 & \epsilon & \epsilon \\
\epsilon &\omega& \omega \\
\epsilon &\omega & \omega\\
\end{array} \right),\label{p413}\\
U^u_L \sim U^u_R \sim U^d_L \sim U_R^e &\sim&  \left(
\begin{array}{ccc}
1 & \epsilon^2 & \epsilon^4 \\
\epsilon^2 &1& \epsilon^2 \\
\epsilon^4 &\epsilon^2 & 1\\
\end{array} \right)\label{p414},
\end{eqnarray}
where $\omega$ is a mixing angle of order one, and coefficients of order one
multiplying $\epsilon$ terms are not exhibited.  These matrices
are of course subject to unitarity constraints.  The CKM mixing matrix for
the quarks is given by $V_{CKM} = (U_L^u)(U_L^d)^\dagger$, which has
small off--diagonal entries as in Eq. (\ref{p414}).\footnote{The Cabibbo angle is
formally of order $\epsilon^2$ from Eq. (\ref{p414}), but coefficients
of order 2 can bring this value to $0.22$ \cite{enkhbat}.}  On the other hand,
the leptonic mixing matrix, $U_{PMNS} = (U_L^e)(U_L^\nu)^\dagger$ will
contain large off--diagonal entries, as in Eq. (\ref{p413}).  This is true
even when $U_L^\nu$, the unitary matrix that diagonalizes the light neutrino
mass matrix is identity.  For $\epsilon \simeq 0.22$, a good fit to all the
mixing angles in the quark and the lepton sector is obtained.\footnote{Small neutrino
masses can be incorporated via the seesaw mechanism by introducing right--handed
neutrinos $\nu_i^c$ with $U(1)$ charges $(1,\,0,\,0)$.  This would lead to
a mild mass hierarchy in the light neutrino sector, as shown in Ref. \cite{enkhbat}.}  Note that
the lopsided nature of $M^d$ and $M^e$ of Eqs. (\ref{Md})-(\ref{Me}) (i.e.,
$(M^d)_{23} \gg (M^d)_{32}$, etc) is crucial
for this result, since large left--handed lepton mixing is correlated with
large right--handed down quark mixing, which however is unobservable in
the SM.

To investigate SUSY flavor violation, we introduce mass insertion parameters defined as
\begin{eqnarray}
(\delta^{d,l}_{LL,RR})_{ij}&=&(U^{\dag d,l}_{L,R},
\tilde{m}^2_{LL,RR}U^{d,l}_{L,R})_{ij}/\tilde{m}^2_{d,l}\label{p411},\\
(\delta^{d,l}_{LR,RL})_{ij}&=&(U^{\dag d,l}_{R,L}
\tilde{m}^2_{LR,RL}U^{d,l}_{L,R})_{ij}/\tilde{m}^2_{d,l}\label{p412},
\end{eqnarray}
where $\tilde{m}^2_{d,l}$ is the average of the diagonal entries of
the scalar mass--squared matrix for the down quarks and charged
leptons and the matrix $\tilde{m}^2_{LR,RL}$ is related to trilinear $A$--terms.
In Table \ref{p4tab6} we list the leading contributions to various
FCNC processes in powers of the small parameter $\epsilon \simeq 0.2$.  
Since the messenger superfields couple with left-handed down quarks
and right-handed charged leptons, the flavor violating
off-diagonal elements are only induced in the quadratic scalar mass
matrices for the left-handed down quarks and right-handed charged
leptons. These matrices are given in Appendix B1. 
The experimental bounds of the mass insertion
parameters $\delta_{LL}$, $\delta_{RR}$ and $\delta_{LR,RL}$ that
are presented in the table were obtained by comparing the hadronic
and leptonic flavor changing processes to their experimental values
\cite{20, Paradisi:2005fk}. We used the branching-ratio expressions
of the decay rates $l_i\rightarrow l_j\gamma$ given in
\cite{Paradisi:2005fk} in order to find the experimental upper
bounds on the leptonic mass insertion parameters that is consistent
with the spectra presented in Table \ref{p4tab4}. The numerical
values of
$\kappa^{d,l}=\frac{m_{b,\tau}A_{d,l}}{\tilde{m}^2_{d,\tau}}$ are
given in Table \ref{p4tab6}. These values are based on the
spectra given in Table \ref{p4tab4}. We can see from Table
\ref{p4tab6} that the $5+\overline{5}$ model is safe from flavor
violation problems as long as $p \geq2$, especially when $r \ll 1$.

\begin{table}[t]
\center
\begin{tabular}{|c|c|c|c|c|c|c|c|c|c|c|} \hline
 SU(5) &$10_1$ &$10_2$&$10_3$& $\overline{5}_1$ &$\overline{5}_2$,$\overline{5}_3$& $5_u$,$\overline{5}_d$&$S$&$10_m$& $\overline{10}_m$& $Z$    \\\hline
 $U(1)$ & 4    & 2    &  0   &      1+p         &             p                   &        0              & -1&  0   &  -$\alpha$           & $\alpha$      \\\hline
\end{tabular}
\caption{The $U(1)$ charge assignments to the $10+\overline{10}$
messenger, MSSM, Z and S
superfields.}\label{p4tab7}\addvspace{0.5cm}
\end{table}

\begin{table}[h!]
\begin{small}
\center
\begin{tabular}{|c|c|c|c|c|c|} \hline
Mass Insertion ($\delta$) & $5+\overline{5}$ & $10+\overline{10}$ &Process&  Exp. Bounds\\
\hline
$(\delta^l_{12})_{LL}$    &  -                 & $\epsilon^{1+2p}$ &  & 0.00028   \\
$(\delta_{12}^l)_{RR}$    & $r$ $\epsilon^{6}$  &         -          &$\mu\rightarrow e\gamma$&   0.0004        \\
$(\delta^l_{12})_{RL,LR}$ & $r$ $\kappa^l_{5}$($\epsilon^{4}$, $\epsilon^{3}$) & $\kappa^l_{10}$ ($\epsilon^{4+2p}$,$\epsilon^{3+2p}$) & &   $1.3\times10^{-6}$ \\
\hline \hline
$(\delta^l_{13})_{LL}$    &        -              &$\epsilon^{1+2p}$& & 0.026\\
$(\delta^l_{13})_{RR}$    &    $r$ $\epsilon^{4}$   &-&$\tau\rightarrow e\gamma$ & 0.04 \\
$(\delta^l_{13})_{RL,LR}$& $r$ $\kappa^l_{5}$($\epsilon^{4}$, $\epsilon^{1}$)  &$\kappa^l_{10}$($\epsilon^{4+2p}$,$\epsilon^{1+2p}$)& & 0.002 \\
\hline\hline
$(\delta^l_{23})_{LL}$    &              -     &$\epsilon^{2p}$ & &   0.02  \\
$(\delta^l_{23})_{RR}$    &   $r$ $\epsilon^{2}$  &-& $\tau\rightarrow \mu\gamma$ & 0.03  \\
$(\delta^l_{23})_{RL,LR}$& $r$ $\kappa^l_{5}$($\epsilon^{2}$,
$1$)
&$\kappa^l_{10}$($\epsilon^{2+2p}$,$\epsilon^{2p}$)& &0.0015\\
\hline \hline
$\left(\sqrt{|\rm{Re}(\delta^d_{12})^2_{LL}|},\sqrt{|\rm{Im}(\delta^d_{12})^2_{LL}}|\right)$ &$\epsilon^6$          &$\epsilon^6$ & & (0.065, 0.0052)  \\
$\left(\sqrt{|\rm{Re}(\delta^d_{12})^2_{RR}|},\sqrt{|\rm{Im}(\delta^d_{12})^2_{RR}|}\right)$ &      -               &$\epsilon^{1+2p}$& & (0.065, 0.0052)  \\
$\left(\sqrt{|\rm{Re}(\delta^d_{12})^2_{LR}|},\sqrt{|\rm{Im}(\delta^d_{12})^2_{LR}}|\right)$ & $\kappa^d_{5} \epsilon^{3}$  &$\kappa^d_{10}$$\epsilon^{3}$& $K-\overline{K}$& (0.007, $5.2\times10^{-5}$)   \\
$\left(\sqrt{|\rm{Re}(\delta^d_{12})^2_{RL}|},\sqrt{|\rm{Im}(\delta^d_{12})^2_{RL}|}\right)$&$\kappa^d_{5}$$\epsilon^{4}$&$\kappa^d_{10}$$\epsilon^{4}$& &
(0.007, $5.2\times10^{-5}$)\\
$\sqrt{|\rm{Re}(\delta^d_{12})_{LL}(\delta^d_{12})_{RR}|}$&-&$\epsilon^{3.5+p}$& &0.00453\\
$\sqrt{|\rm{Im}(\delta^d_{12})_{LL}(\delta^d_{12})_{RR}|}$&-&$\epsilon^{3.5+p}$& &0.00057\\
\hline \hline
$(\rm{Re}\delta^d_{13},\rm{Im}\delta^d_{13})_{LL}$    &  $\epsilon^4$     &$\epsilon^4$& &  (0.238, 0.51)  \\
$(\rm{Re}\delta^d_{13},\rm{Im}\delta^d_{13})_{RR}$    &        -       &$\epsilon^{1+2p}$&$B_d-\overline{B}_d$ & (0.238, 0.51)  \\
$(\rm{Re}\delta^d_{13},\rm{Im}\delta^d_{13})_{LR,RL}$&$\kappa^d_{5}$($\epsilon^{4}$, $\epsilon$) &$\kappa^d_{10}$($\epsilon$,$\epsilon^{4}$)& &(0.0557, 0.125) \\
\hline \hline
$(\delta^d_{23})_{LL}$    &   $\epsilon^2$    &$\epsilon^2$& &   1.19        \\
$(\delta^d_{23})_{RR}$    &          -        &$\epsilon^{2p}$& $B_s-\overline{B}_s$  &   1.19        \\
\hline \hline
$(\delta^d_{23})_{LR,RL}$& $\kappa^d_5$(1,$\epsilon^{2}$) &$\kappa^d_{10}$(1,$\epsilon^{2}$)&$b\rightarrow s\gamma$ & 0.04
\\\hline
\end{tabular}
\caption{The calculated mass insertion parameters for the
$5+\overline{5}$ and $10+\overline{10}$ models and their
experimental upper bounds. The numerical values of $\kappa$'s are
$\kappa^{d}_{5}=0.0045$, $\kappa^{l}_{5}=0.019$,
$\kappa^{d}_{10}=0.002$ and
$\kappa^{l}_{10}=0.0014$. The spectrum corresponds to that of Table. \ref{p4tab4}.}\label{p4tab6}
\end{small}
\end{table}

\subsection{Flavour Violation in $10+\overline{10}$ Model}

The $U(1)$ charge assignments for the messenger, MSSM, $S$, and $Z$
are given in Table \ref{p4tab7}. The superpotential for this model, after field
redefinitions, is
 \begin{eqnarray}
W_{10+\overline{10}}&=&(\lambda'_{u^c}\epsilon^4Q_1+\lambda'_{c^c}\epsilon^2Q_2+\lambda'_{t^c}Q_3)u^c_mH_u+Q_m(\lambda'_{u}\epsilon^4u^c_1+\lambda'_{c}\epsilon^2u^c_2 \nonumber\\
&+&\lambda'_{t}u^c_3)H_u+\lambda'_mQ_mu^c_mH_u+\lambda'_{b}\epsilon^pQ_md^c_3H_d+\lambda'_{\tau}\epsilon^pL_3e^c_mH_d\nonumber\nonumber\\
&+&f_{e^c}\overline{e^c}_me^c_mZ+f_{u^c}\overline{u^c}_mu^c_mZ+f_{Q}\overline{Q}_mQ_mZ.
\end{eqnarray}
 In the $10+\overline{10}$
model, the flavor violating off-diagonal elements are induced in the
scalar matrices of the left-handed down quarks, right-hand down
quarks, and left-handed charged leptons. These matrices are
evaluated in Appendix B2. Using Eqs. (\ref{p411}) and (\ref{p412})
and the unitary transformation given in Eqs. (\ref{p413}) and
(\ref{p414}), the mass insertion parameters for the
$10+\overline{10}$ model are listed in Table \ref{p4tab6}. The
stringent constraint comes from the $\mu\rightarrow e\gamma$ decay
as shown in Table \ref{p4tab6}. The inequality $p\geq1$, or $r\simeq 0$
 should be satisfied in order to suppress the $\mu\rightarrow e\gamma$ decay
process \cite{babupati,rastogi}. 

From Table \ref{p4tab6}, it is clear that all present experimental
limits are satisfied.  Setting the integer $p=1$, we see that the
values close to experimental limits are in CP violation in $K^0$
system, and in $\mu \rightarrow e\gamma$ decay.  The latter is
predicted to occur with an increased experimental sensitivity
of 10 to 100.  Using $\epsilon = 0.22$, we see that new SUSY
contributions to $\epsilon_K$ can be about 30\% of the SM value.
Such new contributions can resolve the apparent discrepancy
between the determinations of $\sin2\beta$ in $B_d$ system
and $\epsilon_K$ \cite{soni}.

\section{Conclusion}

In this paper we have investigated the upper limit on the lightest Higgs boson
mass $m_h$ in gauge mediated supersymmetry breaking models.  In minimal GMSB
models, with all the SUSY particle masses below 2 TeV, the upper limit on
$m_h$ is about 118 GeV.  The vanishing of the trilinear soft term $A_t$
that occurs in minimal GMSB models at the messenger scale sets this restriction on $m_h$, which
could otherwise have been as large as 130 GeV.  We have shown that the
mixing of messenger fields with the MSSM quark and lepton fields can
relax this constraint significantly, primarily because $A_t$ receives
new contributions from the mixed Yukawa couplings at the messenger scale.
Mixing of the messenger fields with the MSSM fields would avoid potential
problems in cosmology with having a stable messenger particle.
We studied two models, one with messengers belonging to $5+\overline{5}$ of
$SU(5)$ unification, and one where they belong to $10+\overline{10}$ of
$SU(5)$.  In the former case, $m_h$ can be as large as about 121 GeV, while
in the latter case $m_h \sim 125$ GeV is realized.  These values of $m_h$
are realized even for $M_{\rm mess} < 10^8$ GeV, which is preferred by cosmology,
since  the gravitino LSP mass would be sub--keV in this case, which avoids
gravitino over-closure of the Universe. The mixed messenger--matter
Yukawa couplings are restricted by the demand that $\tilde{m}^2_{t^c}$ and
$\tilde{m}_{\tau^c}^2$ should not turn negative. We have delineated the
allowed parameter space of these models and have computed the supersymmetric
particle spectrum.  Relatively light stops are realized,
along with $m_h \simeq 125$ GeV, especially in the $10+\overline{10}$ model.

Arbitrary mixing of messenger fields with the MSSM fields can open up
the SUSY flavor problem even in GMSB models.  The increase in $m_h$
and the changes in the SUSY spectrum rely primarily on the mixing of the third family
with the messenger fields.  We have embedded the two models studied here in a unified
framework based on $SU(5)$, along with a flavor $U(1)$ symmetry.  This
$U(1)$ symmetry provides an understanding of the mass and mixing angle
hierarchies in the quark and lepton sectors via the Froggatt--Nielsen mechanism
\cite{fn}.  We have shown that the same $U(1)$ symmetry can prevent bare masses for
the messenger fields, which is necessary for the consistency of gauge mediation.
This $U(1)$ also forbids excessive SUSY flavor violation
by suppressing the mixing of the first two families with the messenger fields.
There could however be residual but small flavor violation arising from the SUSY exchange
diagrams.  We find that new contributions to the CP asymmetry parameter $\epsilon_K$
in the $K$ meson system can be at the $(10-30)\%$ level, which can explain the apparent discrepancy
between $\epsilon_K$ and $\sin 2\beta$ extracted from the $B$ meson system.
We also find that the branching ratio for the decay $\mu \rightarrow e\gamma$
is in the interesting range for next generation experiments.

\section*{Acknowledgments}

We have benefitted from discussions with I. Gogoladze, J. Julio, S. Rai,
Y. Shadmi and especially Z. Chacko.
This work is supported in part by the US Department of Energy Grant No.
DE-FG02-04ER41306.

\section*{Appendix A}
In this Appendix, we present the RGE for the gauge and Yukawa couplings for the two models
considered in the text.

\subsubsection*{A1. RGE for the gauge and Yukawa couplings in the $5+\overline{5}$ messenger model}

Here we present the one--loop RGE for the gauge and Yukawa couplings for the $5+\overline{5}$ model, with the superpotential
given in Eq. (\ref{W5}), valid in the momentum regime  $M_{\rm mess} \leq \mu \leq M_X$. We include the effects of the mixed messenger--matter Yukawa
couplings, and ignore the Yukawa couplings of the first two families.
\begin{eqnarray}
\frac{dg^2_3}{dt}&=&\frac{-g^4_3}{4\pi^2},\nonumber\\
\frac{dg^2_2}{dt}&=&\frac{g^4_2}{4\pi^2},\nonumber\\
\frac{dg^2_1}{dt}&=&\frac{19g^4_1}{20\pi^2},\nonumber\\
\frac{d\lambda^2_t}{dt}&=&\frac{\lambda_t^2}{8\pi^2}\left[6\lambda_t^2+\lambda^{2}_b+\lambda'^{2}_b-\frac{16}{3}g_3^2-3g_2^2-\frac{13}{15}g_1^2\right],\nonumber\\
\frac{d\lambda^2_b}{dt}&=&\frac{\lambda^{2}_b}{8\pi^2}\left[6\lambda^{2}_b+\lambda_{t}^2+\lambda_{\tau}^2+\lambda'^{2}_{\tau^c}+4\lambda'^{2}_b
-\frac{16}{3}g_3^2-3g_2^2-\frac{7}{15}g_1^2\right],\nonumber\\
\frac{d\lambda^2_{\tau}}{dt}&=&\frac{\lambda_{\tau}^2}{8\pi^2}\left[4\lambda_{\tau}^2+3\lambda_{b}^2+3\lambda'^{2}_{\tau^c}+3\lambda'^{2}_b-3g_2^2-
\frac{9}{5}g_1^2\right],\nonumber\\
\frac{d\lambda'^{2}_b}{dt}&=&\frac{\lambda'^{2}_b}{8\pi^2}\left[6\lambda'^{2}_b+4\lambda_{b}^2+\lambda'^{2}_{\tau^c}+\lambda_{t}^2+\lambda_{\tau}^2
+f_d^2-\frac{16}{3}g_3^2-3g_2^2-\frac{7}{15}g_1^2\right],\nonumber\\
\frac{d\lambda'^{2}_{\tau^c}}{dt}&=&\frac{\lambda'^{2}_{\tau^c}}{8\pi^2}\left[4\lambda'^{2}_{\tau^c}+3\lambda_{b}^2+3\lambda'^{2}_b+3\lambda_{\tau}^2
+f_e^2-3g_2^2-\frac{9}{5}g_1^2\right],\nonumber\\
\frac{df_{d}^2}{dt}&=&\frac{f_{d}^2}{8\pi^2}\left[5f_{d}^2+2f_{e}^2+2\lambda'^{2}_b-\frac{16}{3}g_3^2-\frac{4}{15}g_1^2\right],\nonumber\\
\frac{df_{e}^2}{dt}&=&\frac{f_{e}^2}{8\pi^2}\left[4f_{e}^2+3f_{d}^2+\lambda'^{2}_{\tau^c}-3g_2^2-\frac{3}{5}g_1^2\right]~.\nonumber
\end{eqnarray}

\subsection*{A2. RGE for the $10+\overline{10}$ messenger model}

Here we present the one--loop RGE for the various parameters of the $10+\overline{10}$ model, corresponding to the superpotential
given in Eq. (\ref{W10}), valid in the momentum regime  $M_{\rm mess} \leq \mu \leq M_X$.
\begin{eqnarray}
\frac{dg^2_3}{dt}&=&0,\nonumber\\
\frac{dg^2_2}{dt}&=&\frac{g^4_2}{4\pi^2},\nonumber\\
\frac{dg^2_1}{dt}&=&\frac{3g^4_1}{5\pi^2},\nonumber\\
\frac{d\lambda^2_t}{dt}&=&\frac{\lambda_t^2}{8\pi^2}\left[6\lambda_t^2+\lambda^{2}_b+4\lambda'^{2}_{t^c}+5\lambda'^{2}_{t}+3\lambda'^{2}_{m}-\frac{16}{3}g_3^2-3g_2^2
-\frac{13}{15}g_1^2\right],\nonumber\\
\frac{d\lambda^2_b}{dt}&=&\frac{\lambda^{2}_b}{8\pi^2}\left[6\lambda^{2}_b+\lambda_{t}^2+\lambda_{\tau}^2+\lambda'^{2}_{t^c}-\frac{16}{3}g_3^2-3g_2^2
-\frac{7}{15}g_1^2\right],\nonumber\\
\frac{d\lambda^2_{\tau}}{dt}&=&\frac{\lambda_{\tau}^2}{8\pi^2}\left[4\lambda_{\tau}^2+3\lambda_{b}^2-3g_2^2-\frac{9}{5}g_1^2\right],\nonumber\\
\frac{d\lambda'^{2}_{m}}{dt}&=&\frac{\lambda'^{2}_{m}}{8\pi^2}\left[6\lambda'^{2}_{m}+4\lambda'^{2}_{t}+5\lambda'^{2}_{t^c}+3\lambda_{t}^2+f_{Q}^2+f_{u^c}^2
-\frac{16}{3}g_3^2-3g_2^2-\frac{13}{15}g_1^2\right],\nonumber\\
\frac{df_{e^c}^2}{dt}&=&\frac{f_{e^c}^2}{8\pi^2}\left[3f_{e^c}^2+6f_{Q}^2+3f_{u^c}^2-\frac{16}{3}g_3^2-3g_2^2-\frac{12}{5}g_1^2\right],\nonumber\\
\frac{df_{u^c}^2}{dt}&=&\frac{f_{u^c}^2}{8\pi^2}\left[3f_{u^c}^2+6f_{Q}^2+f_{e^c}^2+2\lambda'^{2}_{t^c}+2\lambda'^{2}_{m}-\frac{16}{3}g_3^2-\frac{16}{15}g_1^2\right],\nonumber\\
\frac{df_{Q}^2}{dt}&=&\frac{f_{Q}^2}{8\pi^2}\left[8f_{Q}^2+3f_{u^c}^2+f_{e^c}^2+\lambda'^{2}_{t}+\lambda'^{2}_{m}-\frac{16}{3}g_3^2-3g_2^2-\frac{1}{15}g_1^2\right],\nonumber\\
\frac{d\lambda'^{2}_{t}}{dt}&=&\frac{\lambda'^{2}_{t}}{8\pi^2}\left[6\lambda'^{2}_{t}+3\lambda'^{2}_{t^c}+5\lambda_{t}^2+4\lambda'^{2}_{m}+f_{Q}^2-\frac{16}{3}g_3^2-3g_2^2-\frac{13}{15}g_1^2\right],\nonumber\\
\frac{d\lambda'^{2}_{t^c}}{dt}&=&\frac{\lambda'^{2}_{t^c}}{8\pi^2}\left[6\lambda'^{2}_{t^c}+3\lambda'^{2}_{t}+4\lambda_{t}^2+5\lambda'^{2}_{m}+\lambda_{b}^2+f_{u^c}^2-\frac{16}{3}g_3^2-3g_2^2-\frac{13}{15}g_1^2\right].\nonumber
\end{eqnarray}

\section*{Appendix B}

In this Appendix, we present the new contributions to the scalar masses and the trilinear $A$--terms arising from
messenger--matter mixing in the $5+\overline{5}$ model and the $10+\overline{10}$ model.  We follow the method of Ref. \cite{Chacko}
in our derivations.  The general expressions for the SUSY breaking mass and trilinear parameters, valid in both the $5+\overline{5}$
and the $10+\overline{10}$ model, can be written down as \cite{Chacko}
\begin{eqnarray}
\delta\tilde{m}^2_{Q}(M_{\rm{mess}})&=&-\frac{1}{4}
\left\{\sum_{\lambda}\left(\frac{d\Delta\gamma}{d\lambda}\beta_{>}[\lambda]
-\frac{d\gamma_{<}}{d\lambda}\Delta\beta[\lambda]\right)
\right\}\Lambda^2,\label{p4A1}\\
\delta
\tilde{A}_{abc}(M_{\rm{mess}})&=&\frac{1}{2}\left(\lambda_{a'bc}\Delta\gamma^{a'}_a+\lambda_{ab'c}\Delta\gamma^{b'}_b+\lambda_{abc'}\Delta\gamma^{c'}_c\right)\Lambda
\label{p4A2}~.
\end{eqnarray}
Here the $\lambda$--summation is over the MSSM and mixed MSSM--messenger Yukawa couplings,
$\Delta\beta[\lambda(M_{\rm{mess}})]=\beta_{>}[\lambda(M_{\rm{mess}})]-\beta_{<}[\lambda(M_{\rm{mess}})]$,
and
$\Delta\gamma(M_{{\rm{mess}}})=\gamma_{>}(M_{\rm{mess}})-\gamma_{<}(M_{\rm{mess}})$,
where $\gamma_{>}$($\gamma_{<}$) is the anomalous dimension above
(below) $M_{\rm{mess}}$ and $\beta[\lambda]$ is the beta function
for the Yukawa coupling $\lambda$.  Here $\tilde{A}_{abc}$ is defined through the soft term $V \supset \tilde{A}_{abc} \Phi_a \Phi_b \Phi_c$,
and is related to $A_{abc}$ given in Eqs. (\ref{AB5})- (\ref{AB5pp}) and Eqs. (\ref{AB10})-(\ref{AB10p}) as $\tilde{A}_{abc} = \lambda_{abc} A_{abc}$.

\subsection*{B1.  Soft mass parameters in the $5+\overline{5}$ model}

The (3,3) elements of the
$\Delta\gamma(M_{\rm{mess}})$ matrix for the $Q$, $e^c$ fields, and $\Delta\gamma(M_{\rm{mess}})$ for the $H_d$
field in the $5+\overline{5}$ model are
\begin{eqnarray}
\Delta\gamma_{Q_{33}}(M_{\rm{mess}})&=&-\frac{\lambda'^{2}_b}{8\pi^2},\label{p4A3}\\
\Delta\gamma_{e^c_{33}}(M_{\rm{mess}})&=&-2\frac{\lambda'^{2}_{\tau^c}}{8\pi^2},\\
\Delta\gamma_{H_d}(M_{\rm{mess}})&=&-\frac{3\lambda'^{2}_b+\lambda'^{2}_{\tau^c}}{8\pi^2}.
\end{eqnarray}
The anomalous dimension matrices for the $Q$ and the $e^c$ fields below $M_{\rm{mess}}$ are given by
\begin{eqnarray}
\gamma_{Q_{ij}<}(M_{\rm{mess}})&=&-\frac{1}{8 \pi^2} \left[Y^{u}_{ki}Y_{kj}^{*u}+Y^{d}_{ki} Y_{kj}^{*d}-\frac{8}{3}g_3^2-\frac{3}{2}g_2^2-\frac{1}{30}g_1^2 \right],\label{p417}\\
\gamma_{e^c_{ij}<}(M_{\rm{mess}})&=&-\frac{1}{8\pi^2} \left[2Y_{ik}^eY_{jk}^{*e}-\frac{6}{5}g_1^2 \right]~.
\end{eqnarray}
With the flavor $U(1)$ symmetry, the MSSM Yukawa couplings take the hierarchical form
\begin{eqnarray}
Y^u= \left(
\begin{array}{ccc}
Y^u_{11}\epsilon^8 & Y^u_{12}\epsilon^6 & Y^u_{13}\epsilon^4 \\
Y^u_{21}\epsilon^6  & Y^u_{22}\epsilon^4 & Y^u_{23}\epsilon^2 \\
Y^u_{31}\epsilon^4  & Y^u_{32}\epsilon^2 & Y^u_{33}
\end{array} \right),\\
Y^d=\epsilon^p \left(
\begin{array}{ccc}
Y^d_{11}\epsilon^5 & Y^d_{12}\epsilon^3 & Y^d_{13}\epsilon \\
Y^d_{21}\epsilon^4  & Y^d_{22}\epsilon^2 & Y^d_{23} \\
 Y^d_{31}\epsilon^4  & Y^d_{32}\epsilon^2 & Y^d_{33}
\end{array} \right),\\
Y^e=\epsilon^p \left(
\begin{array}{ccc}
Y^e_{11}\epsilon^5 & Y^e_{12}\epsilon^4& Y^e_{13}\epsilon^4 \\
Y^e_{12}\epsilon^3  & Y^e_{22}\epsilon^2 & Y^e_{23}\epsilon^2 \\
 Y^e_{13}\epsilon & Y^e_{23}& Y^e_{33}
\end{array} \right),
\end{eqnarray}
with $\epsilon \ll 1$, $p=0,1,2$ corresponding to large, medium, and small values of $\tan\beta$, and
all $Y^{u,d,e}_{ij}$ being of order one.
By keeping only the leading $\epsilon^0$ terms we obtain
\begin{eqnarray}
\Delta\beta_{Y^{u}_{i3}}(M_{\rm{mess}})&=&\frac{Y^{u}_{i3}}{16\pi^2}\lambda'^{2}_b,\\
\Delta\beta_{Y^{e}_{ij}}(M_{\rm{mess}})&=&\frac{Y^{e}_{ij}}{16\pi^2}(\lambda'^{2}_{\tau^c}+3\lambda'^{2}_b), \ \ \ \   i\neq 3\\
\Delta\beta_{Y^{e}_{3i}}(M_{\rm{mess}})&=&3\frac{Y^{e}_{3i}}{16\pi^2}(\lambda'^{2}_{\tau^c}+\lambda'^{2}_b),
\end{eqnarray}
with all other contributions suppressed.
The beta-functions for $\lambda'_b$ and $\lambda'_{\tau^c}$ above
$M_{\rm{mess}}$ are given  by
\begin{eqnarray}
\beta_{\lambda'_b>}(M_{\rm{mess}})&=&\frac{\lambda'_{b}}{16\pi^2}\left(6\lambda'^{2}_b+\lambda'^{2}_{e}+(Y^u_{33})^2-\frac{16}{3}g_3^2-3g_2^2
-\frac{7}{15}g_1^2\right),\\
\beta_{\lambda'_{\tau^c}>}(M_{\rm{mess}})&=&\frac{\lambda'_{\tau^c}}{16\pi^2}\left(4\lambda'^{2}_{\tau^c}+3\lambda'^{2}_b-3g_2^2
-\frac{9}{5}g_1^2\right).\label{p4A4}
\end{eqnarray}
Note that $[\gamma_{>},\gamma_{<}]=[\Delta\gamma,\gamma_{<}]$.
Plugging Eqs. (\ref{p4A3})-(\ref{p4A4}) into
Eqs. (\ref{p4A1})-(\ref{p4A2}) and keeping the leading power of
$\epsilon$ we obtain
\begin{eqnarray}
\delta \tilde{m}^2_{e^c}&\sim& \delta\tilde{m}^2_{e^c_3} \left(
\begin{array}{ccc}
\epsilon^{8+2p} & \epsilon^{6+2p}& \epsilon^{4+2p} \\
\epsilon^{6+2p}& \epsilon^{4+2p} & \epsilon^{2+2p} \\
\epsilon^{4+2p} & \epsilon^{2+2p} & 1
\end{array} \right),\\
\delta A_e&\sim& \frac{\Lambda\epsilon^p}{(16\pi^2)}\left(
\begin{array}{ccc}
\epsilon^{5} & \epsilon^{4}& \epsilon^{4} \\
\epsilon^{3}& \epsilon^{2} & \epsilon^{2} \\
\epsilon^{1} & 3(\lambda^{'2}_b+\lambda_{\tau^c}^2) &
3(\lambda^{'2}_b+\lambda_{\tau^c}^2)
\end{array} \right),\\
\delta A_d&\sim&\delta A_b\epsilon^p\left(
\begin{array}{ccc}
\epsilon^{5} & \epsilon^3& \epsilon \\
\epsilon^{4}& \epsilon^{2} & 1 \\
\epsilon^{4} & \epsilon^{2} & 1
\end{array} \right),\\
\delta \tilde{m}^2_{Q}&\sim&\delta\tilde{m}^2_{Q_3}\left(
\begin{array}{ccc}
0 & 0& \epsilon^{4} \\
0& 0 & \epsilon^{2} \\
\epsilon^{4} & \epsilon^{2} & 1
\end{array} \right),\\
\delta \tilde{A}_{t}&=&-\frac{\Lambda}{16\pi^2}Y^u_{33}\lambda'^{2}_b~.
\end{eqnarray}
 Here $\delta\tilde{m}^2_{e^c_3}$ , $\delta\tilde{m}^2_{Q_3}$ and $\delta A_b$ are given
 respectively by Eq.~(\ref{p44}), Eq.~(\ref{p43}) and Eq.~(\ref{p47}).

\section*{B2. Soft mass parameters in the $10+\overline{10}$ model}

From the superpotential $W_{10+\overline{10}}$ of
Eq. (\ref{W10}), we can write $\Delta\gamma_Q$, $\Delta\gamma_{u^c}$
and $\Delta\gamma_{H_u}$ as
\begin{eqnarray}
\Delta\gamma_Q(M_{\rm{mess}})&=&-\frac{\lambda'^{2}_{t^c}}{8\pi^2} ,\\
\Delta\gamma_{u^c}(M_{\rm{mess}})&=&-\frac{2\lambda'^{2}_{t}}{8\pi^2} ,\\
\Delta\gamma_{H_u}(M_{\rm{mess}})&=&-\frac{3}{8\pi^2}\left(\lambda'^{2}_{t}+\lambda'^{2}_{t^c}+\lambda'^{2}_{m}\right).
\end{eqnarray}
The beta-functions for the mixed Yukawa couplings appearing in these
matrices for momenta above $M_{\rm mess}$ are:
\begin{eqnarray}
\beta_{\lambda'_{t^c}>}(M_{\rm{mess}}) &=&\frac{
\lambda'_{t^c}}{16\pi^2}\left(5\lambda'^{2}_{m} + 6\lambda'^{2}_{t^c} +
3\lambda'^{2}_{t}+4(Y^u_{33})^2  \nonumber \right. \\
&-& \left. \frac{16}{3}g_3^2-3g_2^2-\frac{13}{15}g_1^2\right),\\
\beta_{\lambda'_{t}>}(M_{\rm{mess}})&=&\frac{
\lambda'_{t}}{16\pi^2}\left (4\lambda'^{2}_{m} + 6\lambda'^{2}_{t} +
3\lambda'^{2}_{t^c}+5(Y^u_{33})^2\nonumber \right. \\
&-& \left. \frac{16}{3}g_3^2-3g_2^2-\frac{13}{15}g_1^2\right).
\end{eqnarray}

The anomalous dimension matrix $\gamma_{Q<}$ for the $Q$ fields are the same as the ones given in Eq.~(\ref{p417})
for this model.  For the $u^c$ fields, it is given by
\begin{eqnarray}
\gamma_{u^c_{ij}<}(M_{\rm{mess}})&=&-\frac{1}{8\pi^2} \left(2Y_{ik}^uY_{jk}^{*u}-\frac{16}{6}g_3^2-\frac{8}{15}g_1^2\right)~.
 \end{eqnarray}
We also have
\begin{eqnarray}
\Delta\beta_{Y^u_{i3}}(M_{\rm{mess}}) &=&\frac{
Y^u_{i3}}{16\pi^2}\left(3\lambda'^{2}_{m} +
4\lambda'^{2}_{t^c} + 3\lambda'^{2}_{t}\right),  \ \ \ \ i\neq 3 \\
\Delta\beta_{Y^u_{3i}}(M_{\rm{mess}}) &=&
\frac{Y^u_{3i}}{16\pi^2}\left(3\lambda'^{2}_{m} +
3\lambda'^{2}_{t^c} + 5\lambda'^{2}_{t}\right), \ \ \ \ i\neq 3\\
\Delta\beta_{Y^u_{33}}(M_{\rm{mess}}) &=&\frac{
Y^u_{33}}{16\pi^2}\left(3\lambda'^{2}_{m} + 4\lambda'^{2}_{t^c} +
5\lambda'^{2}_{t}\right).
\end{eqnarray}
Using Eqs. (\ref{p4A1}), (\ref{p4A2}) we obtain
\begin{eqnarray}
\delta \tilde{m}^2_{Q}&\sim& \delta \tilde{m}^2_{Q_3}\left(
\begin{array}{ccc}
\epsilon^{8} & \epsilon^{6}& \epsilon^{4} \\
\epsilon^{6}& \epsilon^{4} & \epsilon^{2} \\
\epsilon^{4} & \epsilon^{2} &1
\end{array} \right),\\
\delta \tilde{m}^2_{u^c}&\sim& \delta \tilde{m}^2_{u^c_3}\left(
\begin{array}{ccc}
\epsilon^{8} & \epsilon^{6}& \epsilon^{4} \\
\epsilon^{6}& \epsilon^{4} & \epsilon^{2} \\
\epsilon^{4} & \epsilon^{2} &1
\end{array} \right),\\
\delta \tilde{A}_{u}&\sim& \delta {A_t}\left(
\begin{array}{ccc}
\epsilon^{8} & \epsilon^{6}& \epsilon^{4} \\
\epsilon^{6}& \epsilon^{4} & \epsilon^{2} \\
\epsilon^{4} & \epsilon^{2} & 1
\end{array} \right),\\
\delta \tilde{A}_{d}&\sim& \delta {A_b}\left(
\begin{array}{ccc}
0 & 0& \epsilon^{4} \\
0& 0 & \epsilon^{2} \\
0 & 0 & 1
\end{array} \right),
\end{eqnarray}
where $\delta \tilde{m}^2_{Q_3}$, $\delta \tilde{m}^2_{u^c_3}$, and
$\delta {A_t}$, and $\delta {A_b}$ are given respectively
by Eqs. (\ref{p418}), (\ref{p419}), (\ref{AB10}), (\ref{AB10p}).
Here order one coefficients multiplying each term are to be
understood.

The coupling $\lambda\, \epsilon^p \, \overline{5}\, 10_m\, \overline{5}_d$
induces flavor changing mass terms and trilinear $A$--terms in the $\tilde{d}^c$ and
the $\tilde{e}$ sectors.   These terms are obtained following the same steps as
in the $5+\overline{5}$ model as
\begin{eqnarray}
\delta \tilde{m}^2_{L}&\sim& \delta \tilde{m}^2_{d^c} \sim
\frac{\Lambda^2}{2(16\pi^2)^2}\left(
\begin{array}{ccc}
0 & 0& \epsilon^{1+4p} \\
0& 0 & \epsilon^{4p} \\
\epsilon^{1+4p} & \epsilon^{4p} & \epsilon^{2p}
\end{array} \right),\\
\delta A_{e}&\sim& \frac{\Lambda}{2(16\pi^2)}\left(
\begin{array}{ccc}
\epsilon^{5+2p} & \epsilon^{4+2p}& \epsilon^{4+2p} \\
\epsilon^{3+2p}& \epsilon^{2+2p} & \epsilon^{2+2p} \\
\epsilon^{1+2p} &\epsilon^{2p} &\epsilon^{2p}
\end{array} \right),\\
\delta A_{d}&\sim& \frac{\Lambda}{2(16\pi^2)}\left(
\begin{array}{ccc}
\epsilon^{5+2p} & \epsilon^{3+2p}& \epsilon^{1+2p} \\
\epsilon^{4+2p}& \epsilon^{2+2p} & \epsilon^{2p} \\
\epsilon^{4+2p} &\epsilon^{2+2p} &\epsilon^{2p}
\end{array} \right).
\end{eqnarray}
Here order one couplings multiplying each term are not shown,
but should be understood.

\end{document}